\documentclass[11pt]{article}
\usepackage[a4paper,margin=1in]{geometry}

\usepackage[hidelinks]{hyperref}
\usepackage{orcidlink}
\usepackage{algorithmic}
\usepackage{algorithm}
\usepackage{amsmath,amsfonts}
\usepackage{nomencl}
\usepackage{siunitx}
\usepackage{ifthen}
\usepackage[caption=false,font=normalsize,labelfont=sf,textfont=sf]{subfig}
\usepackage{graphicx}
\usepackage{array}
\usepackage{multirow}
\usepackage{makecell}
\usepackage[T1]{fontenc}
\usepackage{lmodern}
\usepackage{microtype}
\usepackage{geometry}
\usepackage[numbers]{natbib}
\usepackage[utf8]{inputenc}
\usepackage[T1]{fontenc}
\usepackage{xcolor}

\begin{document}

\title{Beyond Backscatter: InSAR coherence from \\ detected SAR images}




\author{
\begin{tabular}{ccc}
Francescopaolo Sica$^{1}$ &
Andrea Pulella$^{1,2}$ &
Michael Schmitt$^{1}$
\end{tabular}
\\[1em]
{\small
$^{1}$Department of Aerospace Engineering, University of the Bundeswehr Munich, Neubiberg, Germany}\\
{\small
$^{2}$Microwaves and Radar Institute, German Aerospace Center (DLR), Weßling, Germany
}
}

\date{}



\maketitle

\begin{abstract}

In this work, we propose a deep learning framework for coherence regression directly from detected SAR images, without the need for accurate coregistration. 
A Residual U-Net is trained using coherence maps derived from precisely coregistered Sentinel-1 SLC data to learn the relationship between backscatter magnitudes and coherence.
The model is trained on 12-day SLC pairs and evaluated across different datasets, including coregistered SLC products and open access analysis-ready data, covering diverse radiometric properties, geometries, and locations.
Experimental results demonstrate that the proposed method achieves high-resolution coherence regression with improved accuracy compared to existing intensity-based approaches. The network generalizes well across diverse geographical locations and even across different temporal baselines that were never seen at training time. Additionally, the ability to operate on globally available analysis-ready data, such as ground range detected data, e.g., distributed through Google Earth Engine, enables its large-scale application in mission design, change monitoring, and diverse mapping tasks. 
\end{abstract}




\section{Introduction}
\label{sec:intro}

Synthetic Aperture Radar (SAR) is a powerful tool for observing the Earth’s surface and inferring its physical properties. SAR data support applications ranging from topographic mapping to land-cover classification and change monitoring. A single SAR acquisition provides information on surface roughness \cite{Engman1987roughness}, spatial texture \cite{Kuplich2003texture, Dekker2003texture}, soil moisture \cite{Hallikainen1985soilmoisture, Dubois1995soilmoisture}, and, to some extent, topography \cite{Recla2024SAR2Height, Xue2022SAR2Height}. When multiple images are acquired with limited spatial or temporal separation, interferometric SAR (InSAR) techniques can be applied to extract the phase difference between coregistered single-look complex (SLC) images, enabling the retrieval of topography and surface motion.

A central quantity in InSAR is the interferometric coherence, defined as the magnitude of the normalized complex correlation between two SLC images. Coherence measures the similarity between acquisitions and depends on temporal and geometric baselines, scattering mechanisms, and system parameters. High coherence ensures reliable phase information, whereas decorrelation directly degrades the quality of interferometric products such as digital elevation models (DEMs) and deformation maps. At the same time, coherence itself is informative about scene dynamics: urban areas typically preserve coherence over time, while vegetation and rapidly evolving surfaces tend to decorrelate.

Beyond interferometric quality assessment, coherence plays an important role in SAR image analysis and mission design. Expected coherence levels guide the selection of revisit time, spatial baseline, and operative frequency to satisfy application-specific requirements. In addition, coherence complements intensity-based features for land-cover classification \cite{Cloude1997PolSAR, Touzi2004Coherence, Strozzi2000LC, Engdahl2003LC, Sica2019RepeatPass, Jacob2020LC, Sica2020paz}, change detection \cite{MontiGuarnieri2018CCD, Stephenson2022DLchangedetection}, and phase unwrapping \cite{Chen2002pu, Zebker1992Phase}. In modern machine and deep learning workflows, coherence further enriches SAR backscatter with contextual information that improves semantic interpretation tasks \cite{Pulella2020, Mazza2019}.

Despite its importance, coherence estimation requires accurately coregistered SLC data, which remains one of the most computationally demanding stages of the interferometric processing chain. In contrast, detected SAR products such as Sentinel-1 Ground Range Detected (GRD) data are widely accessible and routinely employed in large-scale applications relying exclusively on backscatter intensity. Their widespread adoption has been further accelerated by cloud platforms such as Google Earth Engine (GEE), which provide global SAR archives together with scalable processing capabilities.

These considerations motivate the investigation of coherence regression directly from detected SAR backscatter. Intensity-based coherence proxies provide a computationally efficient alternative for rapid data screening, mission analysis, and large-scale monitoring applications without requiring full interferometric processing. Existing approaches include empirical estimators based on backscatter levels \cite{MontiGuarnieri1997} and methods derived from multilook speckle statistics \cite{Aiazzi2003}. However, these techniques typically rely on strong statistical assumptions, require spatial averaging windows that reduce spatial resolution, and still depend on accurately aligned SAR acquisitions.

In this work, we propose a deep learning framework for estimating interferometric coherence directly from detected SAR images. The model is trained using coherence maps derived from precisely coregistered SLC data, while receiving only detected backscatter as input. In this way, the network learns statistical and spatial relationships between paired backscatter observations and their corresponding coherence behavior. 

We further demonstrate that accurate coherence regression is achievable without subpixel coregistration, thereby avoiding one of the most computationally intensive steps of conventional interferometric processing \cite{Sica2026chapter7}.

The main contributions of this work are summarized as follows:
\begin{itemize}
    \item We propose a deep learning framework for high-resolution coherence regression from detected SAR backscatter images.
    
    
    \item We validate the proposed approach across different temporal baselines, geographic regions, polarizations, and operative frequencies.
    
    \item We extend the framework to geocoded Sentinel-1 GRD products from Google Earth Engine, demonstrating its applicability to globally available analysis-ready SAR archives.
\end{itemize}

The remainder of the paper is organized as follows.
Section~\ref{sec:background} introduces the theoretical background of interferometric coherence and reviews related work.
Section~\ref{sec:methodology} presents the proposed framework, including the signal model, preprocessing strategy, network architecture, and training procedure.
Section~\ref{sec:materials} describes the considered datasets.
Section~\ref{sec:experiments} reports the experimental setup and results, including ablation studies, multitemporal analyses, and cross-domain evaluations.
Section~\ref{sec:discussion} discusses the main findings and limitations of the proposed methodology.
Finally, Section~\ref{sec:conclusion} concludes the paper and outlines future research directions.


\section{Background concepts}
\label{sec:background}

Given two complex SAR images $z_1$ and $z_2$ acquired over the same area, the interferogram $\Gamma$ is defined as the complex correlation coefficient:
\begin{equation}
    \Gamma = E\{z_1 z_2^*\} = \sum_\Omega\{z_1 z_2^*\},
    \label{eq:interferogram}
\end{equation}
where $E\{\cdot\}$ denotes the expectation operation over a spatial neighborhood $\Omega$, and $^*$ indicates complex conjugation. The interferometric coherence $\rho$ is the normalized magnitude of $\Gamma$:
\begin{equation}
    \rho = \frac{|\Gamma|}{\sqrt{\sum_\Omega\{|z_1|^2\}\sum_\Omega\{|z_2|^2\}}},
    \label{eq:coherence}
\end{equation}
where $|\cdot|$ denotes the complex magnitude.

Due to the coherent nature of radar imaging, SAR measurements are affected by speckle, which arises from the constructive and destructive interference of multiple scatterers within a resolution cell. In interferometric configurations, coherence measures the similarity between two acquisitions and depends on both acquisition geometry and scene stability. The overall coherence can be expressed as the product of multiple decorrelation factors \cite{Zebker1992Decor, Krieger2007Decor}:
\begin{equation}
\rho = \rho_{\text{SNR}} \cdot \rho_{quant} \cdot \rho_{amb} \cdot \rho_{az} \cdot \rho_{rg} \cdot \rho_{vol} \cdot \rho_{temp},
\label{eq:decorrelation_factors}
\end{equation}
where the terms account for decorrelation due to signal-to-noise ratio, quantization, ambiguities, azimuth and range spectral shifts, volume scattering, and temporal variations.

Coherence is therefore influenced by both acquisition diversity and scattering behavior. Temporal and geometric differences modify the relative phase and scattering conditions between acquisitions, while land cover and scene structure determine the phase stability of the observed targets. Although speckle statistics are partially related to these effects, most decorrelation mechanisms cannot be directly inferred from the amplitude statistics of detected SAR images alone. Consequently, no direct analytical relationship exists between interferometric coherence and the intensity of detected backscatter.

Existing approaches estimate coherence proxies from detected SAR data using local backscatter statistics or multilook speckle models \cite{MontiGuarnieri1997,Aiazzi2003}. However, these methods typically require large averaging windows, leading to reduced spatial resolution and sensitivity to local non-stationarity. Moreover, they still rely on accurately aligned SAR acquisitions and preservation of the original imaging geometry.

Depending on the processing level, SAR data can be represented in different spatial geometries \cite{Sica2026chapter7}. Single-look complex (SLC) products preserve the complex radar signal in slant-range geometry and enable precise interferometric processing through accurate subpixel coregistration. In contrast, Ground Range Detected (GRD) products contain multilooked intensity images projected onto ground-range geometry, where phase information is no longer available and speckle statistics are partially altered by multilooking and resampling operations. Despite these limitations, GRD products are widely adopted in large-scale applications due to their direct georeferencing and accessibility through cloud platforms such as Google Earth Engine (GEE) \cite{GEE}.
Therefore, enabling access to coherence information directly from generic detected SAR datasets, including both SLC-derived intensity products and geocoded GRD archives, would extend the use of coherence-based information to applications where conventional interferometric processing is impractical or unavailable.




\section{Methodology}
\label{sec:methodology}

In this work, we propose a deep learning framework for interferometric coherence regression at full spatial resolution directly from detected SAR backscatter pairs. The approach relies exclusively on radiometrically calibrated backscatter information and does not require subpixel-accurate interferometric coregistration. Coherence maps are generated from precisely coregistered SLC data and used to supervise the training procedure, while the network itself receives only detected backscatter as input.

The proposed signal model is defined as:
\begin{equation}
\hat{\rho} = \mathcal{G}(s_1, s_2),
\label{eq:model}
\end{equation}
where $\hat{\rho}$ denotes the predicted coherence, $\mathcal{G}(\cdot)$ represents the regression model, and $s_1$ and $s_2$ correspond to radiometrically and spatially transformed versions of the backscatter magnitudes $|z_1|$ and $|z_2|$, respectively.

Although interferometric phase information is not directly available to the network, paired backscatter observations still contain statistical and structural cues related to coherence behavior. Temporal stability, texture consistency, scattering mechanisms, and speckle evolution jointly influence the similarity between the two acquisitions. The proposed framework learns these relationships from data, enabling data-driven coherence regression directly from paired detected SAR images.

It is important to emphasize that the proposed methodology does not reconstruct interferometric information explicitly. Rather, it learns a statistical mapping between paired backscatter observations and the expected coherence estimated from the corresponding SLC data. In this sense, the framework can be interpreted as a data-driven coherence regression model trained using automatically generated interferometric supervision.
Furthermore, the proposed framework currently formulates coherence regression as a deterministic prediction problem and does not explicitly model estimator uncertainty or confidence intervals.

\subsection{Data preprocessing}

\begin{figure}[t]
\centering
\includegraphics[width=0.48\textwidth]{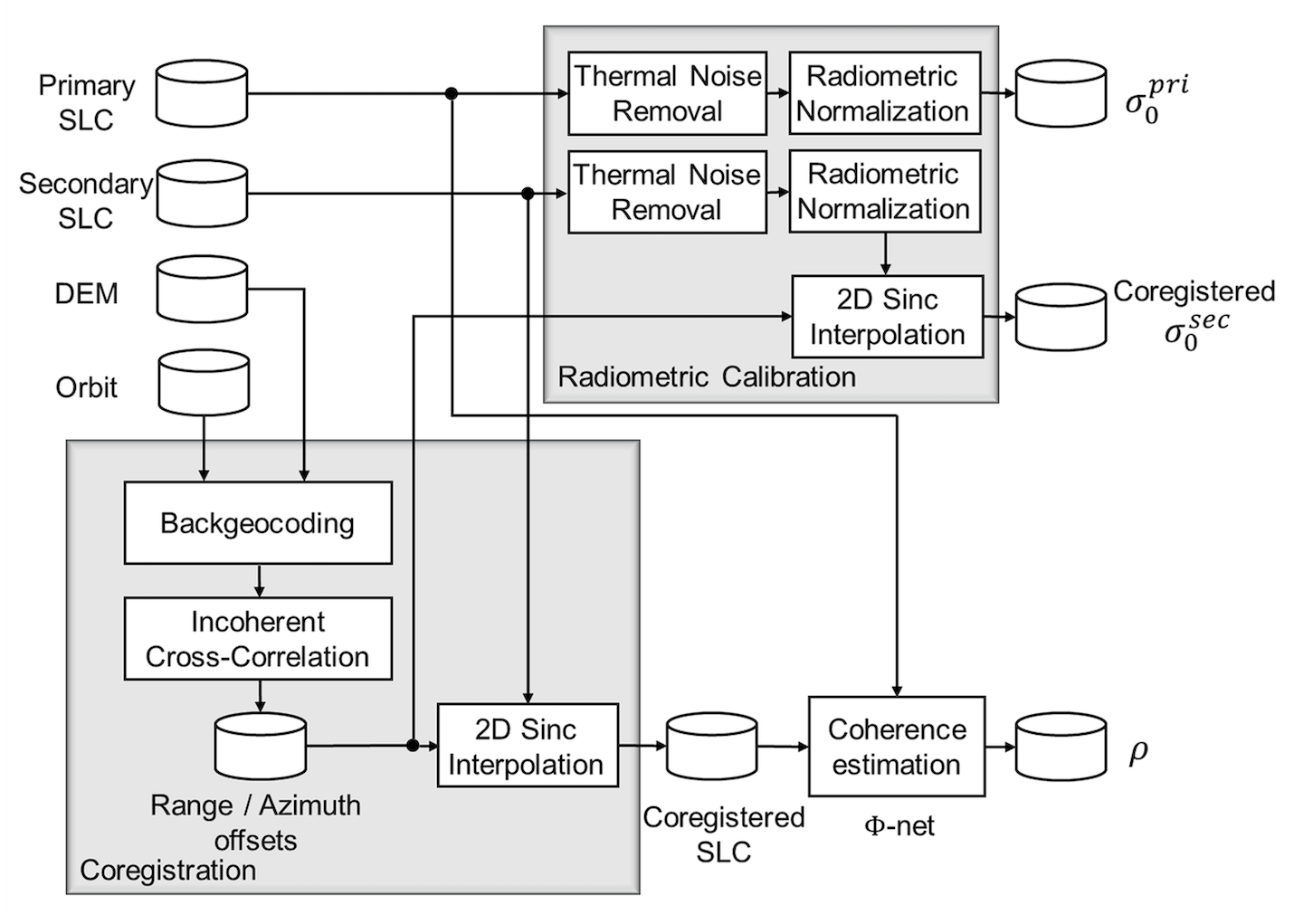}
\caption{Sentinel-1 processing workflow adopted for reference coherence generation from coregistered SLC data.}
\label{fig:flowchart}
\end{figure}

All datasets were radiometrically calibrated in order to convert raw SAR measurements into physically meaningful backscatter coefficients \cite{Small2011}. The calibrated radar backscatter is represented through the sigma nought coefficient ($\sigma^0$), which enables consistent comparison across acquisitions and sensors \cite{Raney1994}. For Sentinel-1 products, radiometric calibration and thermal-noise removal were performed according to the ESA Instrument Processing Facility (IPF) specifications \cite{sentinel1_psd}. The linear sigma nought is computed as:
\begin{equation}
\sigma_0 = \frac{|z|^2 - \eta}{A_\sigma^2},
\label{eq:sigma_nought}
\end{equation}
where $z$ denotes the complex SAR signal, $\eta$ the thermal noise power, and $A_\sigma^2$ the radiometric calibration factor.
The sigma nought values of both acquisitions were converted to decibel scale, clipped to the interval [-20,0]~dB, and linearly normalized to the range [0,1] before being provided as input to the network.
No speckle filtering is applied to the SLC-derived backscatter used for training. The network therefore learns directly from the native speckle statistics of the detected SAR observations.

Reference coherence maps were generated from precisely coregistered Sentinel-1 SLC pairs following the processing workflow illustrated in Figure~\ref{fig:flowchart}. Coregistration was performed using precise orbit information \cite{Fernandez2024poeorb} together with the Copernicus GLO-30 DEM in order to satisfy the stringent geometric accuracy requirements of the TOPS acquisition mode \cite{Yague2016tops}. The final reference coherence was computed using the state-of-the-art $\Phi$-Net framework \cite{Sica2021phinet}, providing high-resolution coherence targets for network training and evaluation.

\subsection{Proposed architecture}
\label{subsec:model}

\begin{figure*}[!t]
    \centering
    \includegraphics[width=0.9\textwidth]{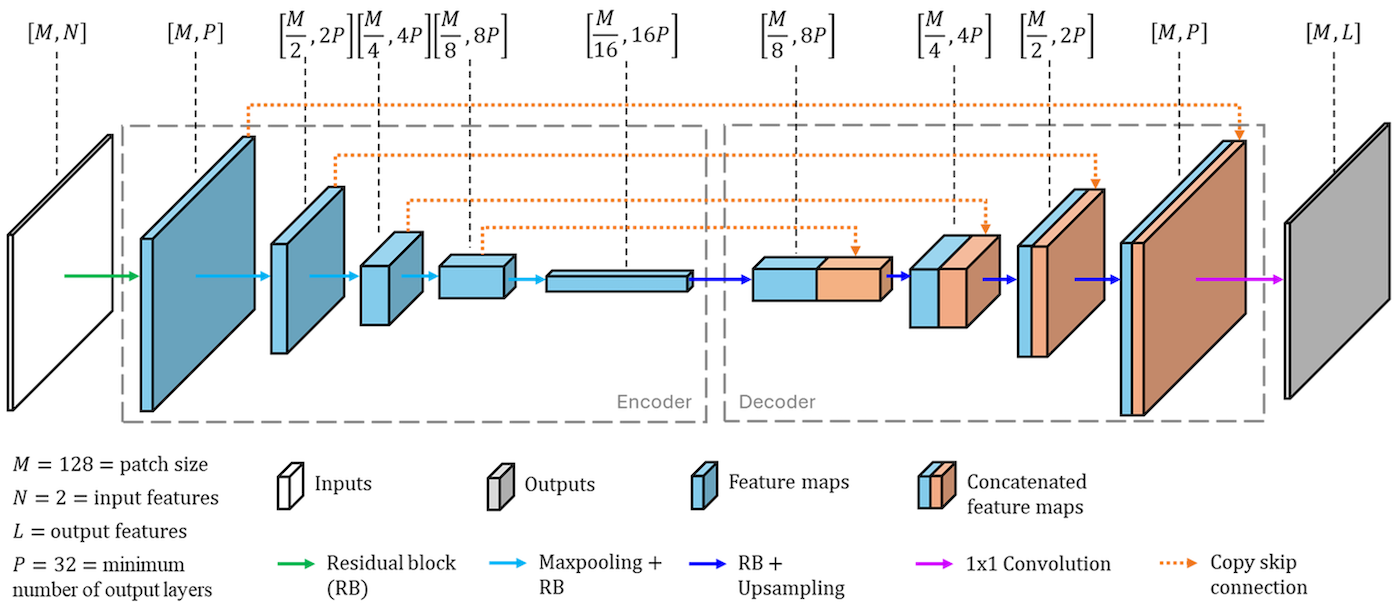}
    \caption{Residual U-Net architecture adopted for coherence regression. The encoder progressively extracts multiscale features, while the decoder reconstructs the coherence map through skip connections and residual blocks.}
    \label{fig:model}
\end{figure*}

The adopted architecture is a Residual U-Net \cite{he2015resunet}, illustrated in Figure~\ref{fig:model}. Both the encoder and decoder consist of four sequential blocks connected by a central bridge. Each encoder level includes a residual block (RB) \cite{Sica2021phinet} followed by a max-pooling operation, while the decoder mirrors this structure by replacing pooling with upsampling at each level. The encoder progressively reduces spatial resolution while increasing feature depth, and the decoder reverses this process. The final layer is a $1 \times 1$ convolution that produces the $L$ output features.
%
The combination of residual and U-Net-style skip connections provides significant advantages in the architecture. 
The former mitigate the vanishing gradient problem and allow for deeper architecture without degradation of performance. The latter bridge encoder and decoder layers at corresponding resolutions, preserving the spatial resolution that might otherwise be lost through downsampling operations.
%
The $N$ input features to the network consist of a normalized representation of the radar backscatter, $\sigma_0$. 
The network is implemented in such a way that it can be trained using either a single input or two inputs. 
The $L$ output features comprise the target's expected value, i.e. the predicted coherence.


\subsection{Training}
\label{subsec:training}

The network is trained exclusively on real SAR data and is performed using the standard approach of stochastic gradient descent.
The model is optimized using a MSE loss function:


%
\begin{equation}
\mathcal{L}(\rho, \hat{\rho}) =
\sum_{k=1}^{K}
\left(\rho_k - \hat{\rho}_k\right)^2
\label{eq:loss_function}
\end{equation}
%

%
%
%

%
\noindent
where \(K\) denotes the total number of pixels considered in the batch, and
\(\rho_k\) and \(\hat{\rho}_k\) represent the reference and predicted values,
respectively, for the \(k\)-th pixel.
All hyperparameters used during training are summarized in Table~\ref{tab:hyperparameters}.
\begin{table}[ht]
    \caption{Hyperparameters}
    \centering
    \renewcommand{\arraystretch}{1.2}
    \setlength{\tabcolsep}{2pt}
    \begin{tabular}{c|cc}
    \hline
    \textbf{Patch size} & \multicolumn{2}{c}{128} \\
    \hline
    \textbf{Batch size} & \multicolumn{2}{c}{64} \\
    \hline
    \textbf{Epochs} & \multicolumn{2}{c}{20} \\
    \hline
    \multirow{4}{*}{\textbf{Learning Rate}} & type & Step-based \\ 
    & initial value & 0.0001 \\ 
    & drop decay & 0.1 \\ 
    & drop rate & 10 \\
    \hline
    \textbf{Optimizer} & \multicolumn{2}{c}{Adam} \\
    \hline
    \textbf{Training perc.} & \multicolumn{2}{c}{90\%} \\
    \hline
    \textbf{Validation perc.} & \multicolumn{2}{c}{10\%} \\
    \hline
    \end{tabular}
    \label{tab:hyperparameters}
\end{table}


\subsection{Fine-tuning strategy for geocoded products}
\label{subsec:fine_tuning}

\begin{figure}[t]
    \centering
    \includegraphics[width=0.48\textwidth]{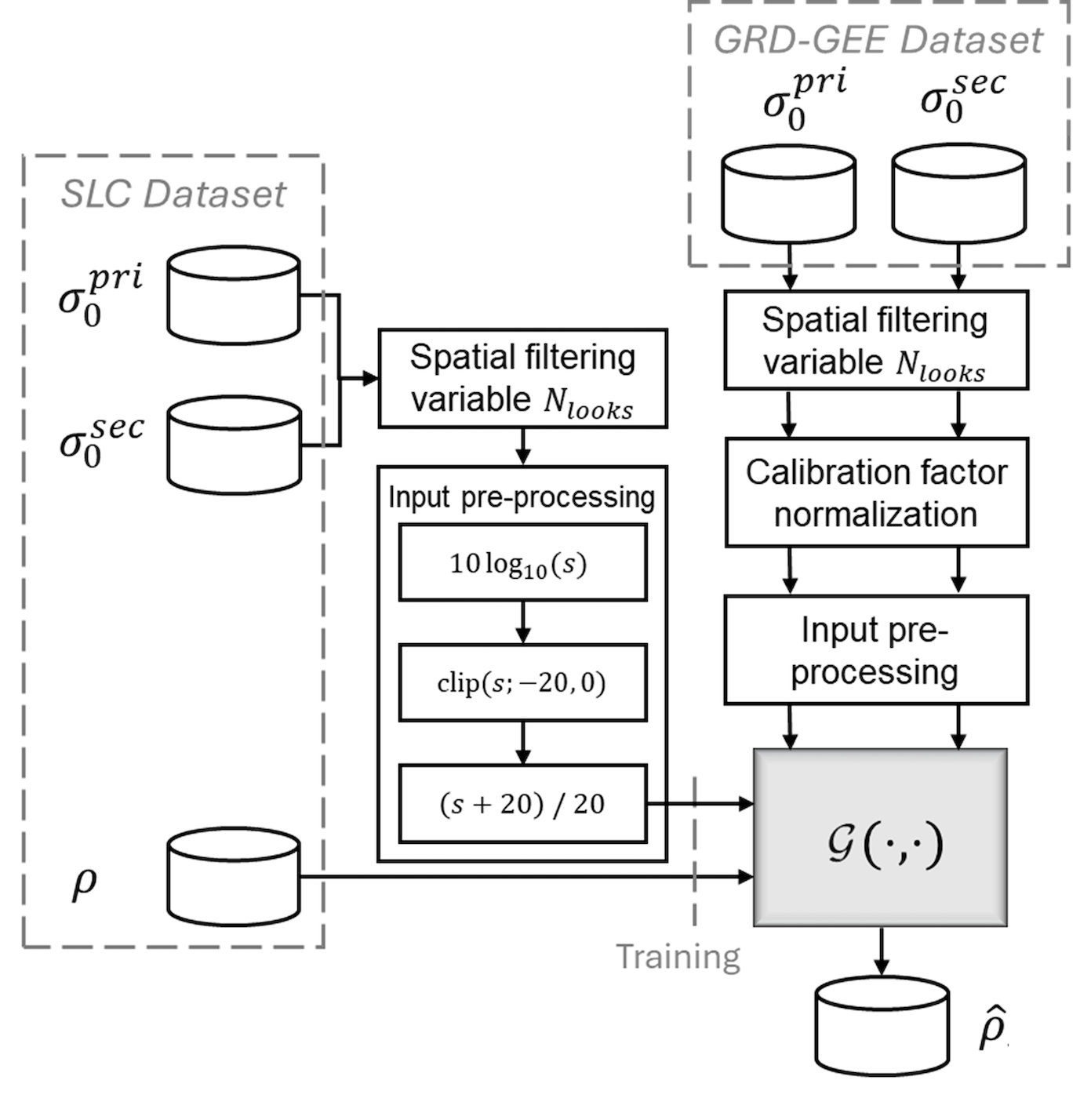}
    \caption{Fine-tuning strategy adopted for adapting the proposed framework from slant-range SLC data to geocoded GRD products. Spatial filtering and radiometric normalization are applied to align the statistical properties of the datasets.}
    \label{fig:geesar_flowchart}
\end{figure}



Compared with slant-range SLC backscatter, geocoded Sentinel-1 GRD products exhibit substantially different statistical and geometric characteristics due to multilooking, terrain correction, spatial interpolation, and reprojection operations performed during the standard preprocessing chain. These transformations modify both the spatial resolution and the speckle statistics of the detected SAR observations, introducing a distribution mismatch with respect to the SLC-derived training data.

To facilitate transfer from interferometric SLC products to analysis-ready GRD data, the proposed framework was fine-tuned using the same SLC Dataset while applying additional preprocessing operations designed to approximate the statistical properties of the GRD products distributed through Google Earth Engine.

First, spatial multilooking with varying numbers of looks was applied during training in order to reproduce the reduced spatial resolution and partially averaged speckle statistics characteristic of GRD products. This operation aims at simulating the smoothing and statistical stabilization effects introduced by multilook processing and geocoding operations.

Second, a radiometric normalization step was introduced to compensate for the systematic radiometric differences observed between the SLC-derived backscatter and the geocoded GRD products. Specifically, the input sigma nought values were rescaled through a global calibration factor estimated from the average radiometric discrepancy between the datasets.

The complete adaptation workflow is summarized in Figure~\ref{fig:geesar_flowchart}. The adopted strategy should be interpreted as a pragmatic domain-adaptation procedure specifically designed for Sentinel-1 GRD products distributed through Google Earth Engine, rather than as a universally optimal calibration framework.
Although the adopted preprocessing parameters were empirically selected for the considered Sentinel-1 products, the obtained results indicate that approximate statistical alignment is sufficient to enable robust transfer across the considered acquisition domains.

\section{Datasets}
\label{sec:materials}

We investigate the problem of coherence regression using different types of SAR data. Specifically, we employ two datasets: (i) the \textit{SLC Dataset} in slant-range geometry , which was created specifically for this study, and 
(ii) the \textit{GEE-SAR Dataset} in geocoded geometry selected from the GEE catalog. The areas of interest are identical across the datasets, while differences arise from the applied spatial interpolation and radiometric characteristics.

\subsection{SLC Dataset}
\label{subsec:slc_description}

This dataset consists of three consecutive Sentinel-1 slices covering areas across Central and Northern Italy. The selection was motivated by the need to capture a wide variety of land-cover types and surface conditions within a compact geographic region. Central and Northern Italy offer substantial heterogeneity, including urban areas, agricultural fields, forests, mountainous terrain, and water bodies.
\textcolor{black}{Additionally, an extended dataset of Sentinel-1 acquisitions, spanning different latitudes, is included exclusively for testing purposes. This set comprises} four consecutive Sentinel-1 acquisitions over the San Francisco Bay area (United States), \textcolor{black}{as well as three 6-day repeat interferometric pairs from the Atacama desert (Chile), the state of Rondonia (Brazil), and the Schwyz alps (Switzerland). A non-conventional Sentinel-1 interferometric pair near the DLR site in Oberpfaffenhofen (Germany), with a relatively large perpendicular baseline of 453 m and a temporal baseline of 84 days, was selected.} 
The selected SLC products are listed in Table~\ref{tab:s1_tds}.

\begin{table}[h]
\caption{Description of the considered Sentinel-1 IW/SLC dataset. From left to right: region of interest (ROI), relative orbit number, sensor name, UTC timestamp, Universally Unique Identifier (UUID), and perpendicular baseline ($B_\perp$) in meters}. 
\label{tab:s1_tds} 
\centering
\renewcommand{\arraystretch}{1.5}
\setlength{\tabcolsep}{2pt}
\begin{tabular}{c c c c c c}
     \hline
     \textbf{ROI} & \textbf{Rel. Orbit} & \textbf{Sensor} & \textbf{Start UTC Time} & \textbf{UUID} & \textbf{$B_\perp$[m]}  \\ 
     \hline
     \multirow{6}{*}{Italy} & \multirow{6}{*}{117} & S1A & 20240704T170606 & DDC8 & - \\ 
     & & S1A & 20240704T170631 & 27CD & - \\ 
     & & S1A & 20240704T170655 & CE80 & - \\ 
     & & S1A & 20240716T170605 & BC48 & 108 \\ 
     & & S1A & 20240716T170630 & 0DE1 & 108 \\ 
     & & S1A & 20240716T170655 & 4258 & 108 \\ 
     \hline
     \multirow{4}{*}{\makecell{San Francisco \\ Bay, U.S.}} & \multirow{4}{*}{35} & S1B & 20200726T020656 & 0D4B & - \\ 
     & & S1A & 20200801T020741 & 115B & 13 \\ 
     & & S1B & 20200807T020656 & 130D & 8 \\ 
     & & S1B & 20200831T020658 & 0F83 & 21 \\ 
     \hline
     \multirow{2}{*}{\makecell{\textcolor{black}{Atacama desert,} \\Chile}} & \multirow{2}{*}{156} & S1B & 20211114T095958 & A4CC & - \\ 
     & & S1A & 20211120T100044 & E386 & 26 \\ 
     \hline
     \multirow{2}{*}{\makecell{\textcolor{black}{Rondonia state,} \\Brazil}} & \multirow{2}{*}{83} & S1A & 20190623T094914 & 4D88 & - \\ 
     & & S1B & 20190629T094837 & 10EE & 56\\ 
     \hline
     \multirow{2}{*}{\makecell{\textcolor{black}{Schwyz alps,} \\Switzerland}} & \multirow{2}{*}{66} & S1B & 20200622T053428 & C4E4 & - \\ 
     & & S1A & 20200628T053501 & 0862 & 108 \\ 
     \hline
     \multirow{2}{*}{\makecell{\textcolor{black}{Oberpfaffenhofen,} \\Germany}} & \multirow{2}{*}{117} & S1A & 20241008T170746 & C8D0 & - \\ 
     & & S1A & 20241231T170742 & 8356 & 453 \\ 
     \hline
\end{tabular}
\end{table}

From the San Francisco Bay area, two of the three available subswaths were selected for analysis, as the first subswath is primarily covered by open ocean and was excluded from further processing. \textcolor{black}{For the remaining four Sentinel-1 test scenarios used in the algorithm evaluation, a dedicated patch of $4096 \times 4096$ pixels was extracted in each case to ensure consistent and comparable analysis across diverse geographic conditions.}
\textcolor{black}{The Oberpfaffenhofen interferometric pair was acquired during the Sentinel-1A orbit control issue reported in \cite{Pinheiro2024}. This choice aims to demonstrate that the proposed methodology is effective even under non-nominal conditions and could potentially be extended to other sensors.}
\textcolor{black}{Finally, to assess the flexibility and robustness of the proposed approach, an additional experiment was conducted using a modified interferometric pair derived from ALOS L-band data, provided within the NISAR sample dataset \cite{nisar_sample_data}. This sample data offers a complementary benchmark for evaluating algorithm performance across frequencies and surface scattering regimes. Also in this case a patch of $4096 \times 4096$ pixels was extracted nearby the Rosamond Dry Lake (United States).}

\subsection{GEE-SAR Dataset}
\label{subsec:grd_gee_description}

Using the same test areas described in Section~\ref{subsec:slc_description}, we additionally retrieved backscatter data from Google Earth Engine (GEE), which provides open-access Sentinel-1 Ground Range Detected (GRD) products under the collection name \textit{COPERNICUS/S1\_GRD} \cite{GEE_GRD_dataset}. This dataset spans from October 3, 2014, to the present and is updated daily, with new assets typically ingested within two days of acquisition. The data are processed into calibrated and ortho-corrected products with multiple spatial resolutions and polarization modes; for this study, the VV polarization at 10~m resolution was selected.

The GEE preprocessing pipeline includes thermal noise removal, radiometric calibration to derive backscatter intensity, and terrain correction using digital elevation models to ensure geometric accuracy and reproject the data into a geographic (latitude–longitude) coordinate system. Finally, the calibrated backscatter values are expressed in decibels (dB) for consistent radiometric representation \cite{GEE_GRD_preprocessing}.

\section{Experimental Results}
\label{sec:experiments}

To evaluate the proposed methodology, the network was tested under a wide range of experimental conditions, including different temporal and spatial baselines, geographic regions, land-cover types, polarizations, operative frequencies, and geometric representations. The experiments are designed to assess both the accuracy of the proposed coherence regression framework and its robustness across acquisition conditions and data domains. Quantitative performance is evaluated using the root-mean-square error (RMSE), and comparisons with state-of-the-art coherence estimation approaches from detected SAR backscatter are provided.

All training procedures are performed using exclusively 12-day Sentinel-1 interferometric pairs. Nevertheless, the evaluation is intentionally conducted under different configurations in order to investigate the temporal and spatial generalization capability of the proposed model.

The experimental analysis is organized as follows.

Section~\ref{subsec:slc_12days} first evaluates the proposed framework on a standard 12-day Sentinel-1 interferometric pair over the San Francisco Bay area in slant-range geometry. An ablation study is additionally performed by training the same architecture using only the primary acquisition as input, allowing the contribution of temporal information to be explicitly assessed.

Section~\ref{subsec:slc_multi_analysis} investigates temporal generalization by evaluating the network on 0-, 6-, 12-, and 36-day temporal baselines over areas characterized by different scattering behaviors, including urban, forested, and cultivated regions.

The robustness of the proposed methodology across acquisition conditions is further evaluated through experiments involving different geographic regions and biomes, cross-polarized Sentinel-1 data, large perpendicular baselines, and different operative frequencies.

Finally, the methodology is extended to geocoded Sentinel-1 GRD products from the \textit{GEE-SAR Dataset}. In this setting, the fine-tuned model is evaluated across multiple temporal baselines to assess its transferability to analysis-ready SAR products characterized by different radiometric and geometric properties.


\subsection{Baseline coherence regression on Sentinel-1 SLC data}
\label{subsec:slc_12days}

\begin{figure*}[!t]
    \centering
    \includegraphics[width=0.9\textwidth]{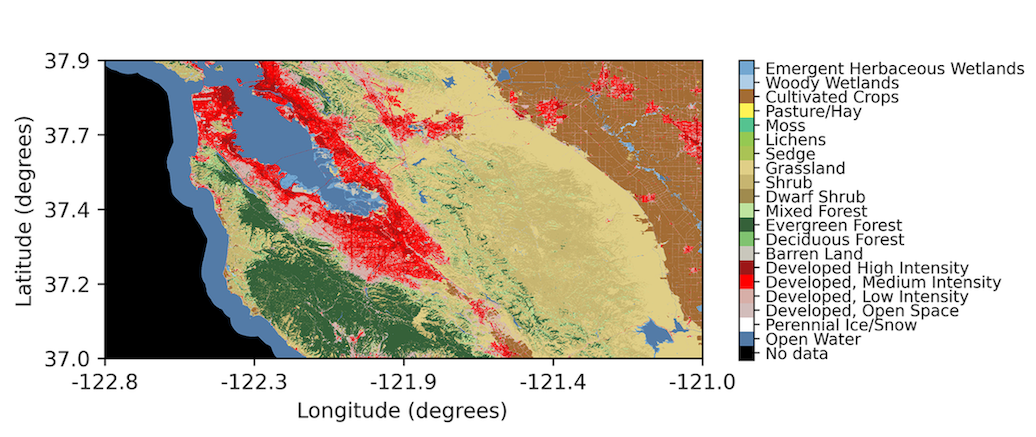}
    \caption{NLCD land-cover map over the San Francisco Bay area, United States, used as a reference for interpreting coherence regression results across different land-cover types.}
    \label{fig:nlcd_map}
\end{figure*}

\begin{figure*}[!t]
    \centering
    \includegraphics[width=0.95\textwidth]{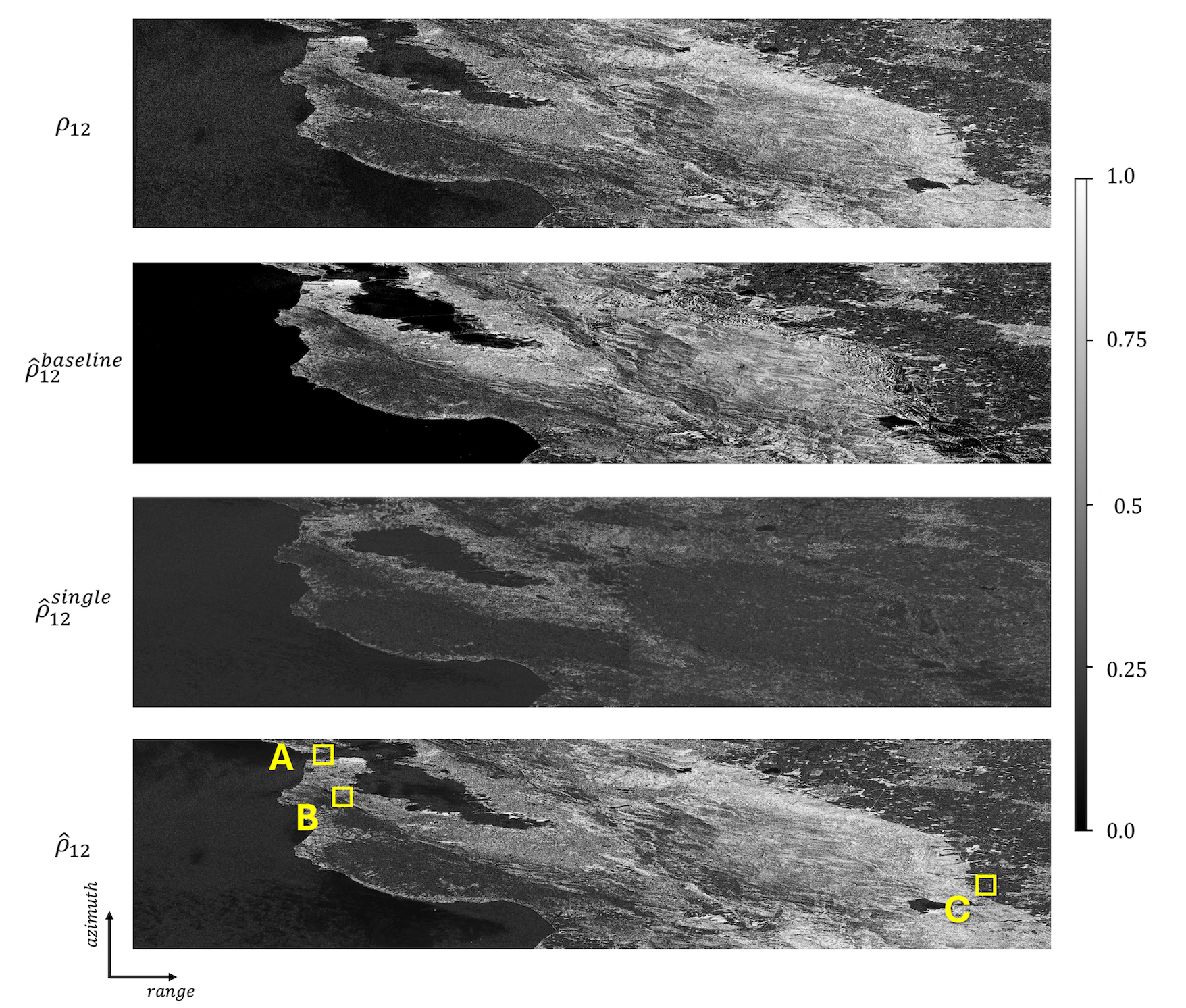}
    \caption{Coherence regression results obtained on a 12-day Sentinel-1 interferometric pair over the San Francisco Bay area in slant-range geometry. From top to bottom: primary and secondary backscatter images, reference coherence, and coherence predicted by the proposed framework. Yellow squares indicate the representative crops used for local-scale analyses.}
    \label{fig:sanfrancisco_12days}
\end{figure*}

\begin{figure*}[!t]
    \centering
    \includegraphics[width=0.7\textwidth]{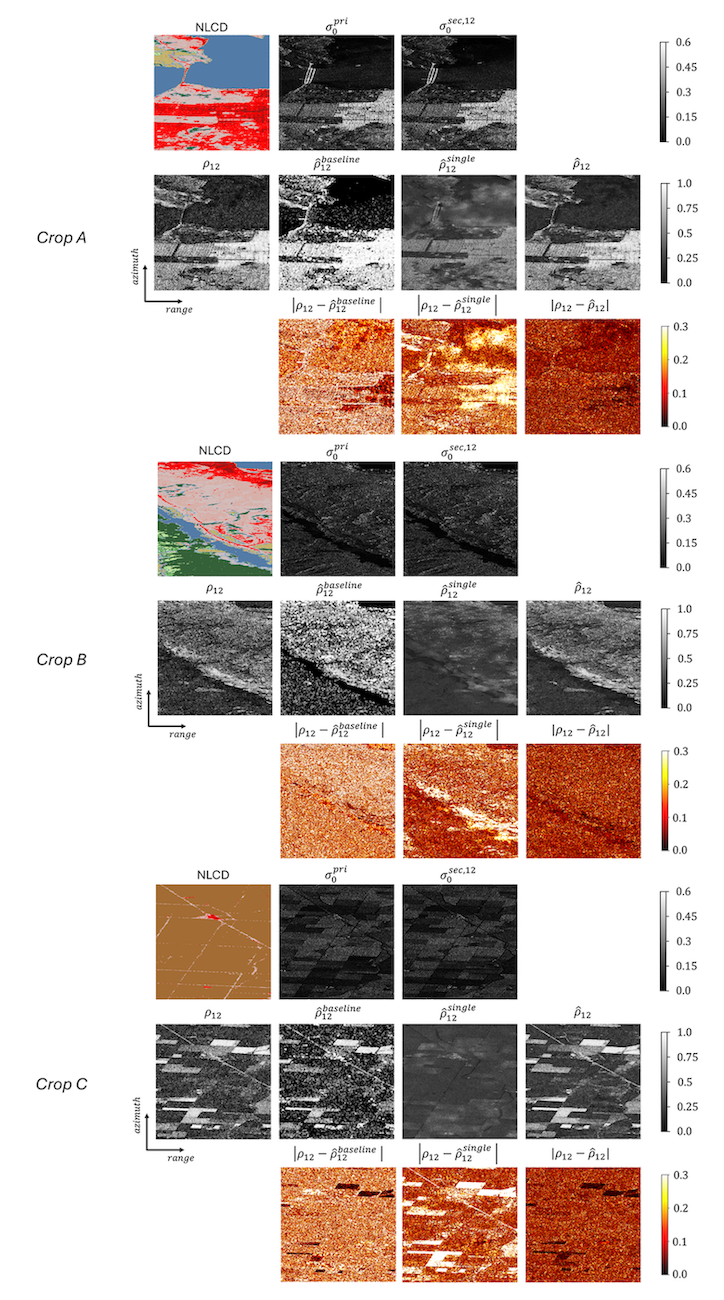}
    \caption{Three selected crops, A, B, and C, over the San Francisco Bay area are shown, with each crop represented in three rows. From left to right: the first row presents the NLCD map, the primary and secondary $\sigma_0$ images, respectively. The second row shows the reference, baseline \cite{MontiGuarnieri1997}, single, and estimated coherence maps, respectively. The third row displays the corresponding absolute error maps for the baseline, single, and estimated coherence maps, respectively.}
    \label{fig:sanfrancisco_12days_crops}
\end{figure*}

This experiment evaluates the proposed framework on a standard 12-day Sentinel-1 interferometric pair acquired over the San Francisco Bay area. Although the model was trained exclusively on Sentinel-1 data over Italy, the test scenario involves a geographically independent region characterized by different land-cover distributions and scattering conditions. This experiment therefore provides an initial assessment of the spatial generalization capability of the proposed methodology.

For scene interpretation, the National Land Cover Database (NLCD) provided by the United States Geological Survey (USGS) \cite{Homer2015} is used as reference land-cover information. The considered area includes heterogeneous scattering conditions, including urban environments, agricultural fields, wetlands, forests, and water bodies. The corresponding NLCD map is shown in Figure~\ref{fig:nlcd_map}.

Figure~\ref{fig:sanfrancisco_12days} reports the coherence regression results obtained on the full San Francisco Bay scene. The predicted coherence maps are in close agreement with the reference coherence derived from the SLC data, while preserving fine spatial structures and sharp transitions between different land-cover types. Urban regions maintain high coherence levels, whereas vegetation and water bodies exhibit stronger decorrelation, consistently with the expected interferometric behavior.

To further investigate the role of temporal information, an ablation study was conducted by retraining the same network architecture using only the primary acquisition as input. This experiment aims to verify whether the network learns meaningful relationships between paired SAR observations or merely exploits correlations between backscatter amplitude and coherence. The single-input configuration resulted in a substantial performance degradation, confirming that the proposed framework effectively exploits the statistical relationship between the two acquisitions.

For a more detailed local-scale analysis, Figure~\ref{fig:sanfrancisco_12days_crops} shows three representative crops characterized by different scattering mechanisms and land-cover conditions. The proposed method preserves detailed spatial structures while reducing the smoothing artifacts and residual speckle typically observed in conventional intensity-based coherence estimators.

Quantitative results are summarized in Table~\ref{tab:rmse_12days}. The proposed framework consistently outperforms the baseline coherence estimator \cite{MontiGuarnieri1997} across all evaluated regions, achieving lower RMSE values while preserving higher spatial detail.

\begin{table}[h]
\caption{RMSE obtained on the 12-day Sentinel-1 coherence regression experiment over the San Francisco Bay area. Results are reported for the baseline estimator \cite{MontiGuarnieri1997}, the proposed dual-input framework ($\hat{\rho}$), and the single-input ablation model ($\hat{\rho}^{single}$).}
\label{tab:rmse_12days} 
\centering
\renewcommand{\arraystretch}{1.5}
\begin{tabular}{c|ccc}
    \hline
    \multicolumn{4}{c}{\textbf{RMSE}} \\
    \hline
    \textbf{Area} & $\hat{\rho}^{baseline}$ & $\hat{\rho}^{single}$ & $\hat{\rho}$ \\
    \hline
    \textbf{Overall} & $0.275$ & $0.261$ & $0.108$ \\ 
    \hline
    \textbf{Crop A} & $0.280$ & $0.295$ & $0.108$ \\ 
    \hline
    \textbf{Crop B} & $0.288$ & $0.257$ & $0.118$ \\ 
    \hline
    \textbf{Crop C} & $0.242$ & $0.265$ & $0.117$ \\ 
    
    \hline
\end{tabular}
\end{table}

\subsection{Generalization across interferometric temporal baselines}
\label{subsec:slc_multi_analysis}

\begin{figure*}[!t]
    \centering
    \includegraphics[width=0.7\textwidth]{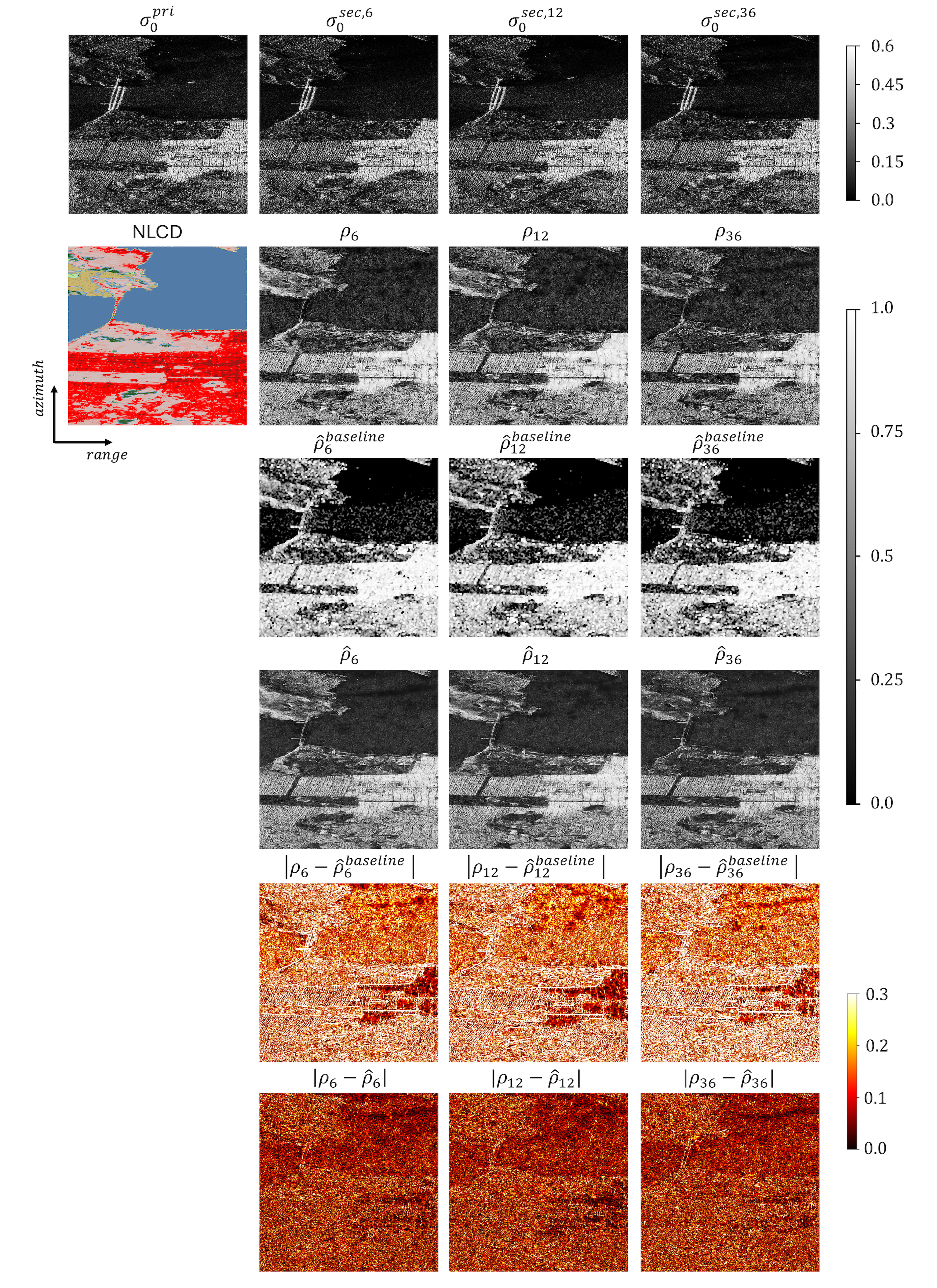}
    \caption{Temporal coherence analysis over \textit{Crop A}. From top to bottom: input backscatter images, reference coherence maps, coherence estimated using the baseline approach \cite{MontiGuarnieri1997}, coherence predicted by the proposed framework, and absolute prediction error for 6-, 12-, and 36-day temporal baselines.}
    \label{fig:cropA_analysis}
\end{figure*}

\begin{figure*}[!t]
    \centering
    \includegraphics[width=0.7\textwidth]{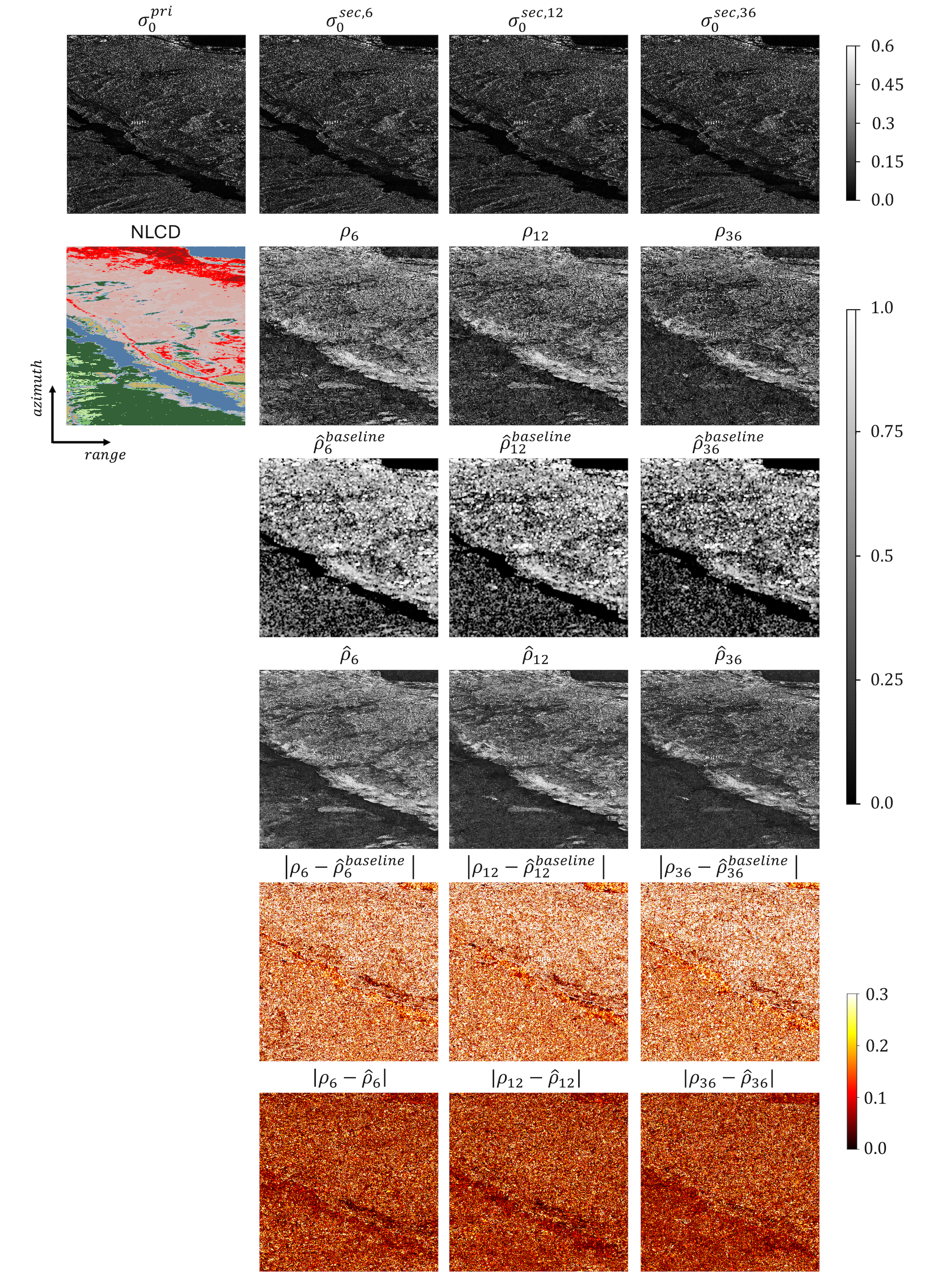}
    \caption{Temporal coherence analysis over Crop B using the same visualization layout as in Figure~7.}
    \label{fig:cropB_analysis}
\end{figure*}

\begin{figure*}[!t]
    \centering
    \includegraphics[width=0.7\textwidth]{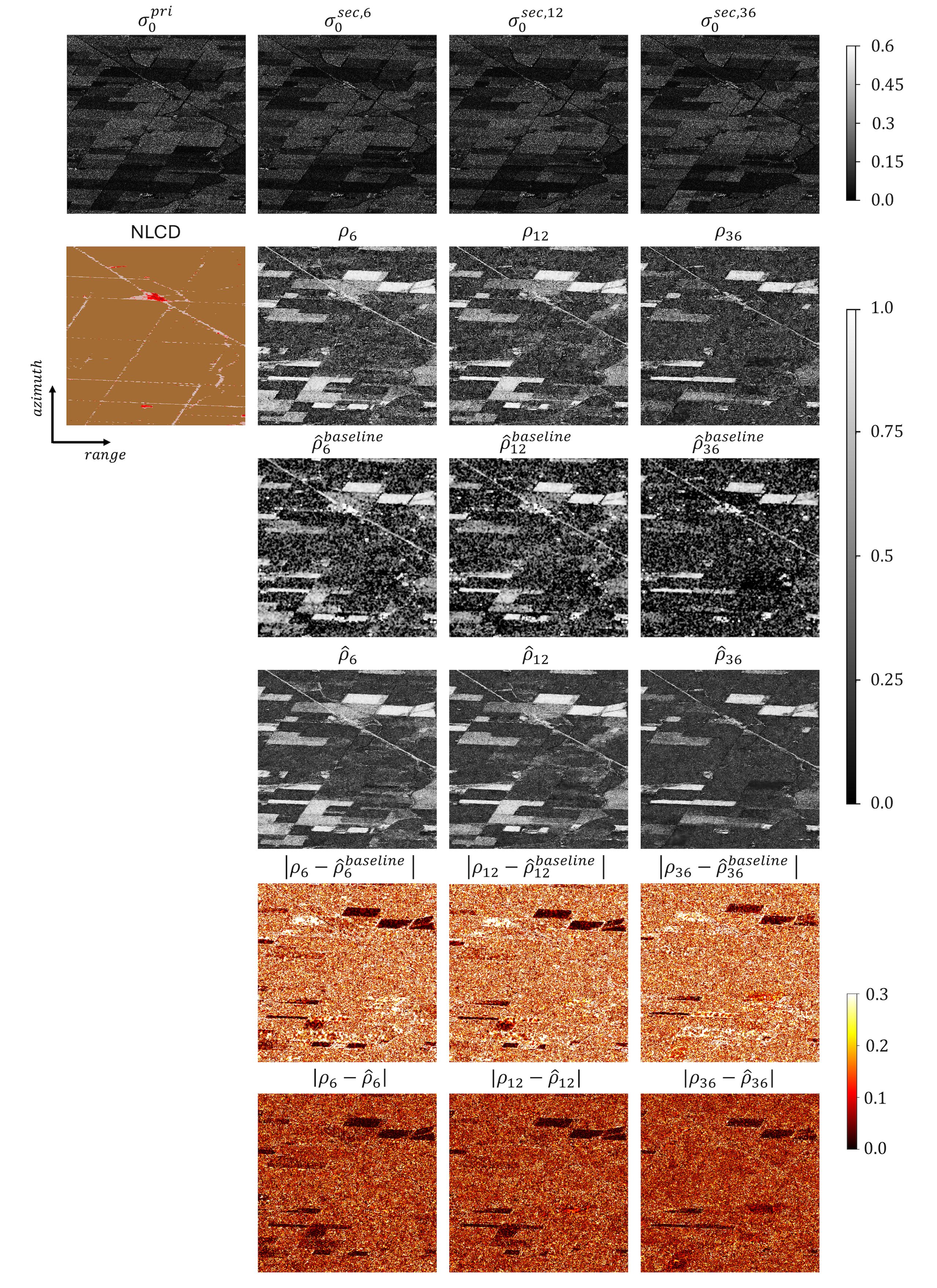}
    \caption{Temporal coherence analysis over Crop C using the same visualization layout as in Figure~7.}
    \label{fig:cropC_analysis}
\end{figure*}

\begin{figure*}[!t]
    \centering
    \includegraphics[height=0.6\textwidth]{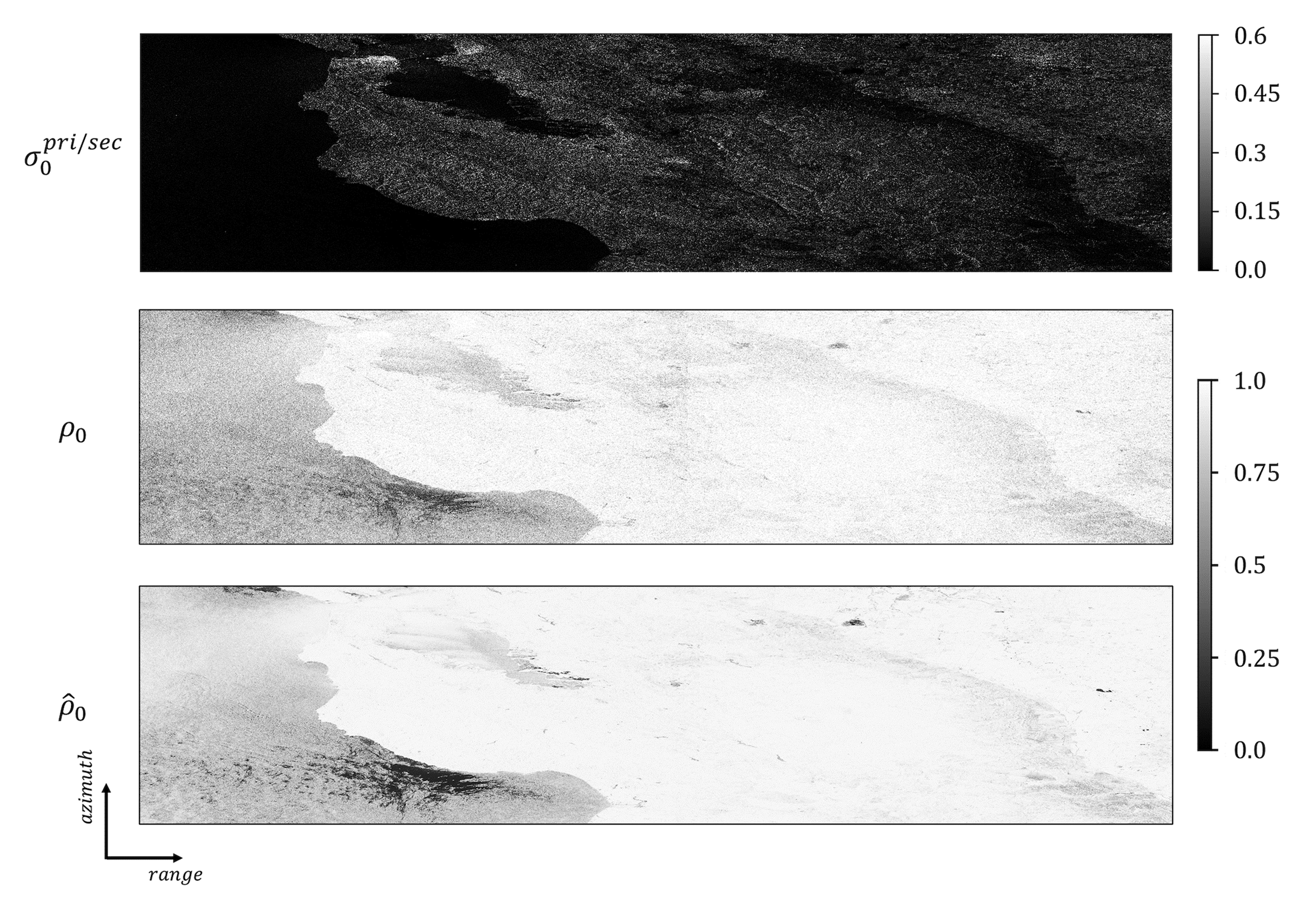} 
    \caption{Temporal coherence analysis at zero baseline. From top to bottom: input backscatter image, SNR-limited reference coherence, and coherence predicted by the proposed framework.}
    \label{fig:sanfrancisco_00days}
\end{figure*}

To assess temporal generalization, the proposed framework was evaluated on interferometric pairs characterized by temporal baselines different from those used during training. Although the network was trained exclusively on 12-day Sentinel-1 pairs, inference was performed on 0-, 6-, 12-, and 36-day configurations by fixing the same primary acquisition and varying the secondary image. All experiments were conducted in slant-range geometry.

The analysis focuses on three representative $1024 \times 1024$ crops, denoted as \textit{Crop~A}, \textit{Crop~B}, and \textit{Crop~C}, highlighted in Figure~\ref{fig:sanfrancisco_12days}. These regions are characterized by substantially different scattering behaviors and land-cover conditions, including urban environments, vegetation, water bodies, and cultivated areas.

Figure~\ref{fig:cropA_analysis} reports the temporal analysis over \textit{Crop~A}, which is primarily dominated by impervious surfaces and open water. The predicted coherence maps remain in close agreement with the reference coherence across all temporal baselines, accurately preserving stable urban structures while reproducing the moderate decorrelation observed over water surfaces. In contrast, the baseline estimator \cite{MontiGuarnieri1997} exhibits stronger smoothing artifacts and residual speckle fluctuations, particularly over low-coherence regions.

The results over \textit{Crop~B}, shown in Figure~\ref{fig:cropB_analysis}, further demonstrate the capability of the proposed framework to capture heterogeneous temporal decorrelation behavior. In this area, vegetation regions progressively decorrelate with increasing temporal baseline due to volume scattering and temporal variability, whereas impervious surfaces preserve higher coherence levels. The proposed model successfully reproduces these spatial trends while maintaining finer structural details than the baseline estimator.

Figure~\ref{fig:cropC_analysis} presents the temporal analysis over \textit{Crop~C}, mainly characterized by cultivated fields. Despite the strong temporal decorrelation associated with agricultural areas, the predicted coherence maps remain spatially consistent and preserve detailed field structures. Compared with the baseline method, the proposed approach produces sharper coherence transitions and significantly reduces the propagation of speckle artifacts.

An additional experiment was conducted in the zero-baseline configuration by using identical backscatter images as both primary and secondary inputs. In this scenario, coherence is exclusively limited by the signal-to-noise ratio (SNR), since temporal and geometric decorrelation terms vanish. Figure~\ref{fig:sanfrancisco_00days} shows that the predicted coherence closely follows the expected SNR-limited behavior, demonstrating consistency with the expected SNR-driven coherence behavior.

For the zero-baseline case, the reference coherence was computed as \cite{Zebker1992Decor}:
\begin{equation}
    \rho_0 = \frac{1}{1+\text{SNR}^{-1}} ,
    \label{eq:rho_snr}
\end{equation}
where the SNR is estimated as:
\begin{equation}
    \text{SNR} = \frac{|z|^2}{P_{noise}} ,
    \label{eq:snr}
\end{equation}
with $z$ denoting the complex SAR signal and $P_{noise}$ the estimated noise power. Low-coherence regions observed over open water areas are mainly associated with reduced SNR conditions.

Quantitative results are summarized in Table~\ref{tab:rmse_multitemp}. Despite being trained exclusively on 12-day interferometric pairs, the proposed framework generalizes successfully across all evaluated temporal baselines, consistently outperforming the baseline estimator \cite{MontiGuarnieri1997}. The RMSE values remain relatively stable across the different configurations, indicating that the network learns generalized coherence-related statistical relationships rather than memorizing a specific temporal baseline.

\begin{table}[h]
\caption{Performance results on the multitemporal coherence predictions. The table displays the root mean square error (RMSE) at 0-, 6-, 12-, and 36-day when applying the selected baseline algorithm \cite{MontiGuarnieri1997} and the proposed CNN on the overall area and on crops A, B and C.} 
\label{tab:rmse_multitemp} 
\centering
\renewcommand{\arraystretch}{1.5}
\begin{tabular}{c|c|cc}
    \hline
    \multicolumn{3}{c}{\textbf{RMSE on the outputs}} \\
    \hline
    \textbf{Area} & \textbf{Temporal baseline} & \cite{MontiGuarnieri1997} & Ours \\
    \hline
    \multirow{4}{*}{\textbf{Overall}} & 0-day & $0.257$ & $0.068$ \\
    & 6-day & $0.269$ & $0.106$ \\
    & 12-day & $0.275$ & $0.108$ \\ 
    & 36-day & $0.283$ & $0.116$ \\
    \hline
    \multirow{4}{*}{\textbf{Crop A}} & 0-day & $0.202$ & $0.070$ \\
    & 6-day & $0.277$ & $0.107$ \\
    & 12-day & $0.280$ & $0.108$ \\ 
    & 36-day & $0.295$ & $0.111$ \\
    \hline
    \multirow{4}{*}{\textbf{Crop B}} & 0-day & $0.200$ & $0.066$ \\
    & 6-day & $0.280$ & $0.119$ \\
    & 12-day & $0.288$ & $0.118$ \\ 
    & 36-day & $0.296$ & $0.118$ \\
    \hline
    \multirow{4}{*}{\textbf{Crop C}} & 0-day & $0.085$ & $0.059$ \\
    & 6-day & $0.244$ & $0.119$ \\
    & 12-day & $0.242$ & $0.117$ \\ 
    & 36-day & $0.241$ & $0.113$ \\
    \hline
\end{tabular}
\end{table}

\subsection{Generalization across acquisition conditions}
\label{subsec:other_acq_conditions}

The previous experiments evaluated the proposed framework under standard Sentinel-1 interferometric configurations and varying temporal baselines. In the following analyses, we further investigate the robustness of the model under progressively more challenging acquisition conditions, including different polarization modes, geographic regions and biomes, large perpendicular baselines, and different operative frequencies. These experiments are designed to assess the transferability of the learned representations beyond the acquisition conditions observed during training.

\subsubsection{Generalization across polarizations}
\label{subsec:vh_tuning}

\begin{figure*}[!t]
    \centering
    \includegraphics[width=0.95\textwidth]{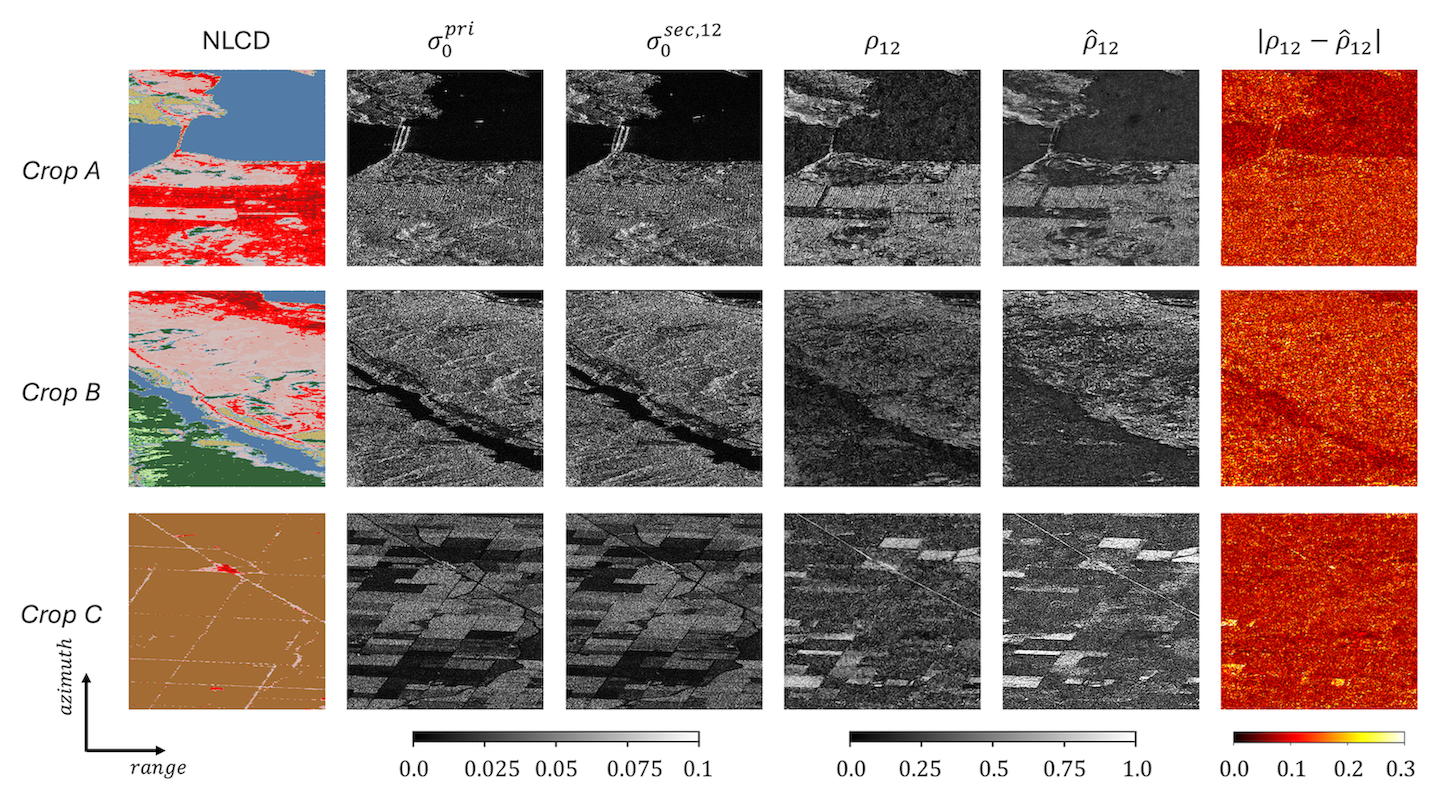}
    \caption{Coherence regression results on VH-polarized Sentinel-1 data over the three representative crops. For each crop, the figure reports the reference coherence, the coherence predicted by the proposed framework, and the corresponding absolute error.}
    \label{fig:vh_analysis}
\end{figure*}

To assess the robustness of the proposed framework across polarization modes, the model trained exclusively on VV-polarized Sentinel-1 data was evaluated on VH interferometric pairs. Since VH backscatter typically exhibits lower intensity levels than VV, a simple radiometric normalization was applied in order to reduce the statistical mismatch between the training and inference distributions. Specifically, the VH sigma nought values were rescaled using a global calibration factor derived from the average ratio between VV and VH backscatter intensities.

Despite the absence of VH data during training, the proposed framework successfully generalized to cross-polarized acquisitions. Figure~\ref{fig:vh_analysis} shows the coherence regression results over the same representative crops considered in the previous experiments. The predicted coherence maps remain spatially consistent with the reference coherence and preserve the main decorrelation patterns across urban, vegetation, and cultivated areas.

Quantitative results are summarized in Table~\ref{tab:rmse_acq_cond}. Although a moderate performance degradation is observed with respect to the VV experiments, the obtained RMSE values confirm that the proposed framework maintains robust coherence prediction capability across different polarization configurations. These results suggest that the network learns generalized statistical relationships associated with coherence behavior rather than polarization-specific radiometric patterns alone.

\subsubsection{Spatial generalization across geographic regions and biomes}
\label{subsec:other_scenarios}

\begin{figure*}[!t]
    \centering
    \includegraphics[width=\textwidth]{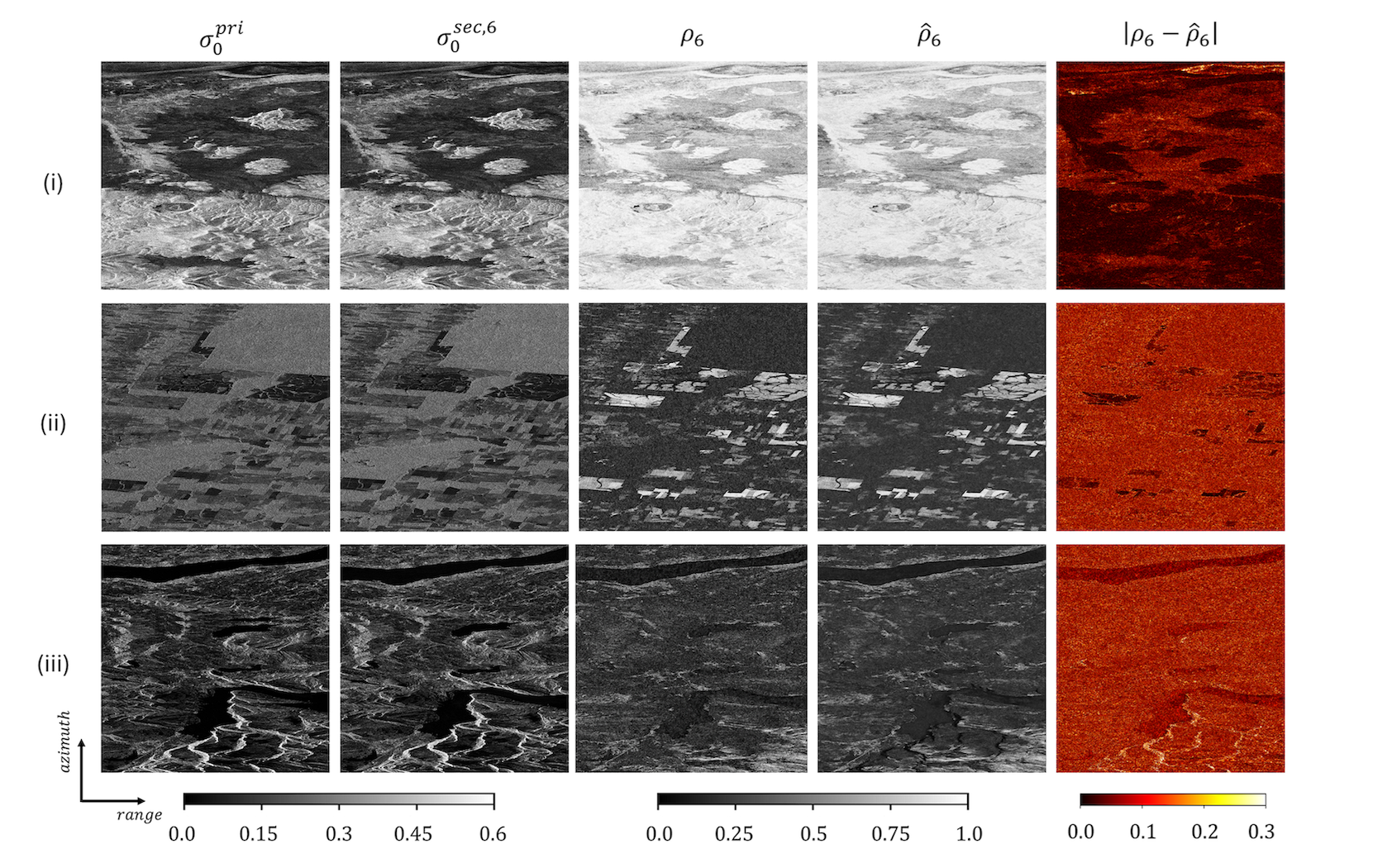}
    \caption{Coherence regression results over three geographically distinct Sentinel-1 scenarios characterized by different biome and scattering conditions. For each scene, the figure reports the input backscatter images, the reference coherence, the coherence predicted by the proposed framework, and the corresponding absolute prediction error.}
    \label{fig:vv_others_analysis}
\end{figure*}

The proposed framework was further evaluated across geographically distinct regions characterized by substantially different environmental conditions, scattering mechanisms, and coherence regimes. The objective of this experiment is to assess the spatial generalization capability of the network beyond the distributions observed during training.

Figure~\ref{fig:vv_others_analysis} summarizes the results obtained over three representative scenarios. Scene~(i), located in the Atacama Desert, corresponds to an arid and sparsely vegetated environment characterized by highly stable scattering conditions and consistently high coherence levels. Scene~(ii), located in Rondonia State, Brazil, contains dense tropical forest areas interspersed with cleared land, producing strong spatial variations in temporal decorrelation due to volume scattering. Scene~(iii), located in the Schwyz Alps, Switzerland, is characterized by mountainous topography and rapidly varying terrain-induced scattering conditions.

Across all considered scenarios, the proposed framework successfully captures the main coherence spatial patterns and adapts to different decorrelation behaviors. In particular, the network accurately distinguishes stable and decorrelating regions in the tropical forest scene, while preserving coherent structures in arid and mountainous environments. Minor underestimation effects are mainly observed in the steepest alpine regions, where geometric distortions and local scattering variability become more pronounced.

Quantitative results are reported in Table~\ref{tab:rmse_acq_cond}. The obtained RMSE values confirm the robustness of the proposed methodology across diverse geographic regions and biome-dependent scattering conditions, despite the network being trained on a limited set of Sentinel-1 acquisitions over Italy.

\subsubsection{Generalization across large perpendicular baselines}
\label{subsec:large_bperp}

\begin{figure*}[!t]
    \centering
    \includegraphics[width=0.95\textwidth]{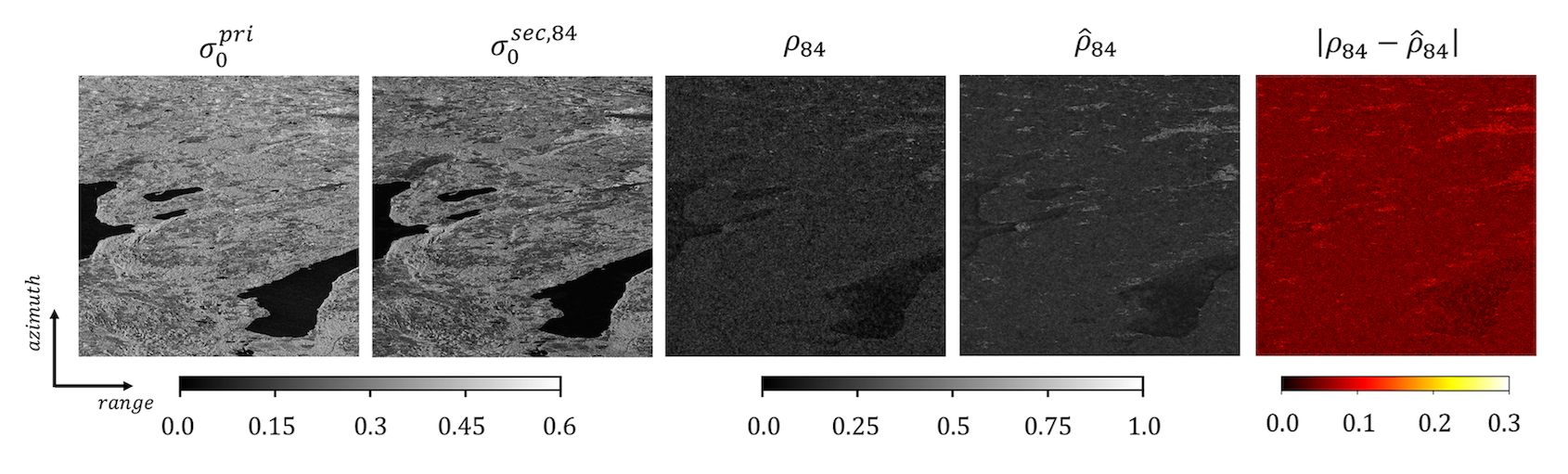}
    \caption{Coherence regression results obtained under a large-baseline interferometric configuration over the Oberpfaffenhofen region (Scene~(iv)). The figure reports the input backscatter images, the reference coherence, the coherence predicted by the proposed framework, and the corresponding absolute prediction error.}
    \label{fig:op_vv_crop}
\end{figure*}

The proposed framework was additionally evaluated under a challenging large-baseline interferometric configuration. In this experiment, a Sentinel-1 interferometric pair acquired over the Oberpfaffenhofen region (Scene~(iv)) near Munich was considered, characterized by a perpendicular baseline of approximately 453 m and a temporal separation of 84 days. These acquisition conditions introduce stronger geometric sensitivity together with increased temporal and volumetric decorrelation effects.

Figure~\ref{fig:op_vv_crop} shows the coherence regression results obtained over a $4096 \times 4096$ patch covering heterogeneous land-cover conditions, including water bodies, urban infrastructure, and forested regions. Despite the significantly different acquisition geometry with respect to the training data, the proposed framework successfully reproduces the main coherence structures across the scene. Stable man-made structures preserve high coherence levels, whereas forested areas exhibit stronger decorrelation due to volume scattering and temporal variability.

As expected, the increased acquisition diversity leads to a moderate performance degradation compared with the standard Sentinel-1 experiments, yielding an RMSE of 0.126 (Table~\ref{tab:rmse_acq_cond}). Nevertheless, the obtained results demonstrate that the proposed methodology remains robust even under acquisition conditions substantially different from those observed during training.

\subsubsection{Generalization across operative frequencies}
\label{subsec:frequency_tuning}

\begin{figure*}[!t]
    \centering
    \includegraphics[width=0.95\textwidth]{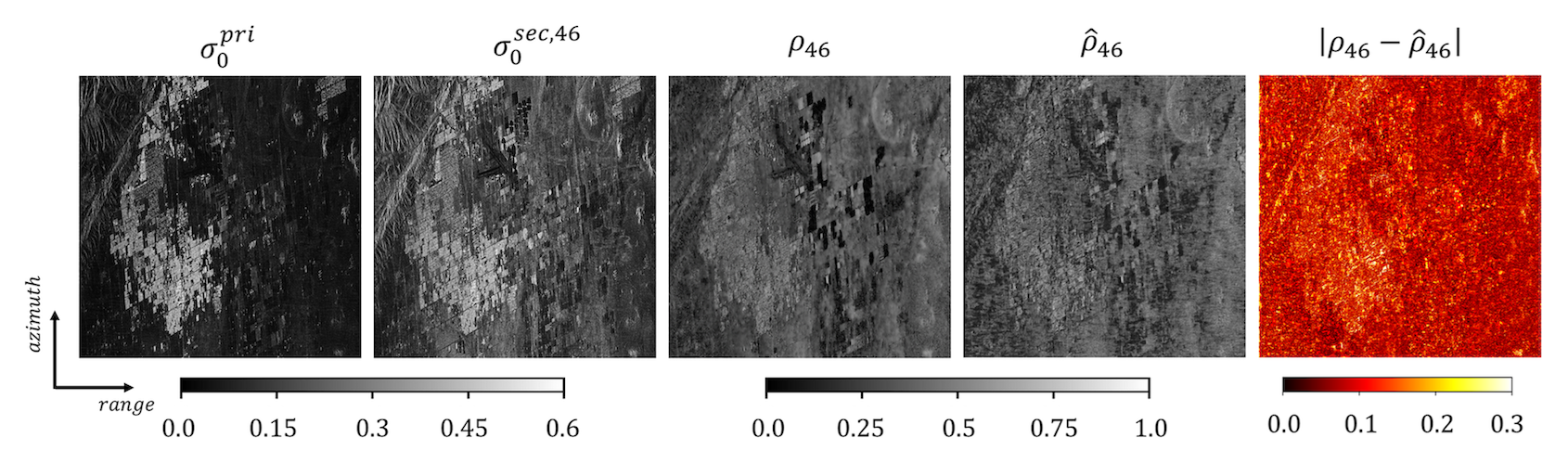}
    \caption{Coherence regression results obtained on an ALOS L-band HH-polarized interferometric pair over the Rosamond Dry Lake region (Scene~(v)). The figure reports the input backscatter images, the reference coherence, the coherence predicted by the proposed framework, and the corresponding absolute prediction error.} 
    \label{fig:alos_Lband_hh}
\end{figure*}

To further evaluate the robustness of the proposed framework across sensor configurations and wavelength regimes, the model was tested on ALOS L-band SAR data acquired over a rural area near Rosamond Dry Lake, California (Scene~(v)). This experiment represents a substantially different operating condition with respect to the Sentinel-1 training data, involving a different sensor, operative frequency, polarization mode (HH), and interferometric geometry.

Figure~\ref{fig:alos_Lband_hh} shows the coherence regression results obtained on the considered L-band interferometric pair, characterized by a perpendicular baseline of approximately 300~m. Despite the absence of L-band data during training, the proposed framework successfully reproduces the main spatial coherence structures and decorrelation patterns across the scene.

As expected, the domain shift between Sentinel-1 C-band VV data and ALOS L-band HH acquisitions leads to a moderate performance degradation compared with the previous experiments, yielding an RMSE of 0.215 (Table~\ref{tab:rmse_acq_cond}). Nevertheless, the obtained results demonstrate that the proposed framework retains meaningful coherence prediction capability even under different frequency, polarization, and geometries.

\begin{table*}[h]
\caption{RMSE obtained for the generalization experiments on different acquisition conditions of Section \ref{subsec:other_acq_conditions}.}
\label{tab:rmse_acq_cond} 
\centering
\renewcommand{\arraystretch}{1.5}
\begin{tabular}{c|c|c|c|c|c|c|c}
    \hline
    \multicolumn{8}{c}{\textbf{RMSE - Different acquisition scenarios}} \\
    \hline
    \multicolumn{2}{c}{\textbf{Polarization}} & \multicolumn{2}{c}{\textbf{Geographic}} & \multicolumn{2}{c}{\textbf{Spatial baseline}} & \multicolumn{2}{c}{\textbf{Frequency}} \\
    \hline
    \textbf{Area} & Ours & \textbf{Area} & Ours & \textbf{Area} & Ours & \textbf{Area} & Ours \\
    \hline
    \textbf{Crop A} & $0.143$ & \textbf{Scene (i)} & $0.055$ & \textbf{Scene (iv)} & $0.126$ & \textbf{Scene (v)} & $0.215$ \\
    \hline
    \textbf{Crop B} & $0.150$ & \textbf{Scene (ii)} & $0.105$ \\
    \hline
    \textbf{Crop C} & $0.141$ & \textbf{Scene (iii)} & $0.139$ \\
    \hline
\end{tabular}
\end{table*}


\subsection{Transfer to geocoded GRD products}
\label{subsec:gee_tuning}

\begin{figure*}[!t]
    \centering
    \includegraphics[width=0.75\textwidth]{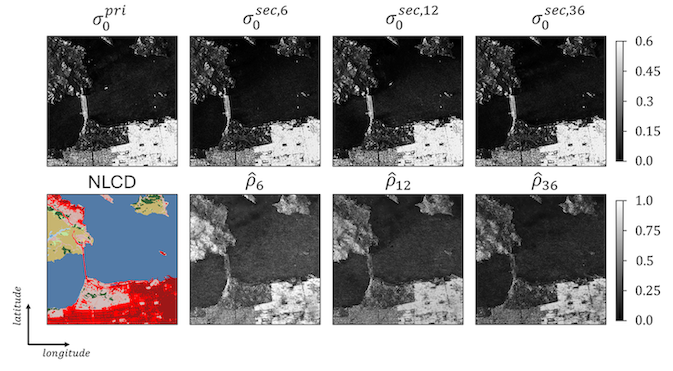}
    \caption{Temporal coherence predictions obtained on geocoded Sentinel-1 GRD data over Crop A.}
    \label{fig:sanfrancisco_GEE_cropA}
\end{figure*}

\begin{figure*}[!t]
    \centering
    \includegraphics[width=0.75\textwidth]{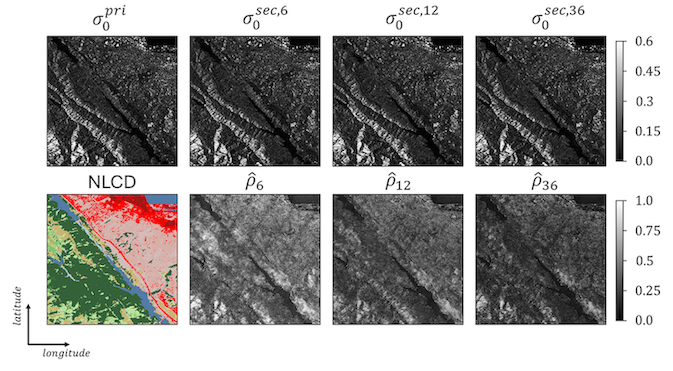}
    \caption{Temporal coherence predictions obtained on geocoded Sentinel-1 GRD data over Crop B.}
    \label{fig:sanfrancisco_GEE_cropB}
\end{figure*}

\begin{figure*}[!t]
    \centering
    \includegraphics[width=0.75\textwidth]{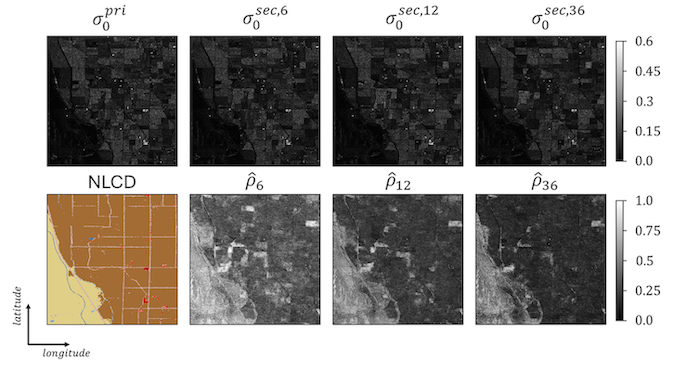}
    \caption{Temporal coherence predictions obtained on geocoded Sentinel-1 GRD data over Crop C.}
    \label{fig:sanfrancisco_GEE_cropC}
\end{figure*}

To evaluate the applicability of the proposed framework to operational analysis-ready SAR products, the methodology was extended to geocoded Sentinel-1 GRD data from the \textit{GEE-SAR Dataset}. Compared with the slant-range SLC experiments, this scenario introduces different radiometric and geometric characteristics due to multilooking, terrain correction, and geocoding operations. Consequently, a dedicated fine-tuning strategy was adopted in order to adapt the statistical properties of the input backscatter to those of the GRD products.

For consistency with the previous experiments, the same representative regions analyzed in Section~\ref{subsec:slc_multi_analysis} were selected in geocoded geometry. Figures~\ref{fig:sanfrancisco_GEE_cropA}--\ref{fig:sanfrancisco_GEE_cropC} report the temporal coherence predictions obtained over the three selected crops for multiple temporal baselines.

Figure~\ref{fig:sanfrancisco_GEE_cropA} shows the results over \textit{Crop~A}, mainly characterized by urban and open-water areas. Despite the different spatial sampling and radiometric properties of the GRD products, the predicted coherence maps remain globally consistent with the coherence behavior observed in the corresponding slant-range experiments.

Figure~\ref{fig:sanfrancisco_GEE_cropB} reports the results over \textit{Crop~B}, which contains heterogeneous vegetation and urban regions. The predicted coherence maps correctly reproduce the temporal decorrelation observed over forested areas, although some residual stripe-like artifacts are visible in the geocoded products. These structures are likely associated with internal resampling operations and geometric interpolation effects present in the \texttt{COPERNICUS/S1\_GRD} processing chain.

Finally, Figure~\ref{fig:sanfrancisco_GEE_cropC} shows the results over an agricultural region dominated by cultivated fields. Similarly to the slant-range experiments, the proposed framework successfully captures the different temporal decorrelation behaviors across neighboring fields and grassland areas.

Overall, these experiments demonstrate that the proposed methodology can be successfully transferred from interferometric SLC data to globally available geocoded GRD products. This capability is particularly relevant for large-scale and cloud-based SAR applications, where analysis-ready products distributed through platforms such as Google Earth Engine are substantially more accessible than precisely coregistered SLC interferometric stacks.


\section{Discussion}
\label{sec:discussion}

\begin{figure}[!t]
    \centering
    \includegraphics[height=0.25\textwidth]{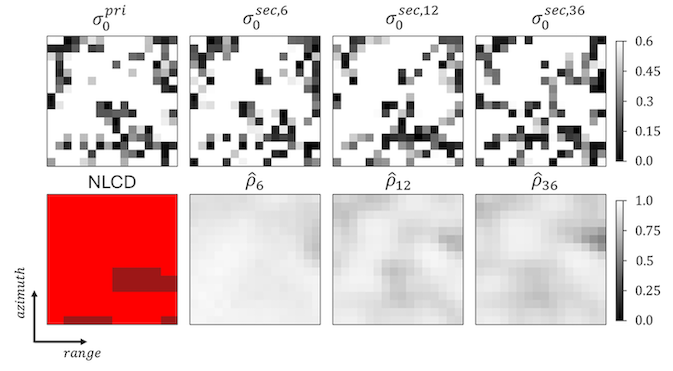}
    \caption{Local-scale analysis over a developed urban region (Crop I). The figure reports the temporal evolution of the input backscatter and the coherence predicted by the proposed framework across different temporal baselines.}
    \label{fig:crop16x16_impervious}
\end{figure}


\begin{figure}[!t]
    \centering
    \includegraphics[height=0.25\textwidth]{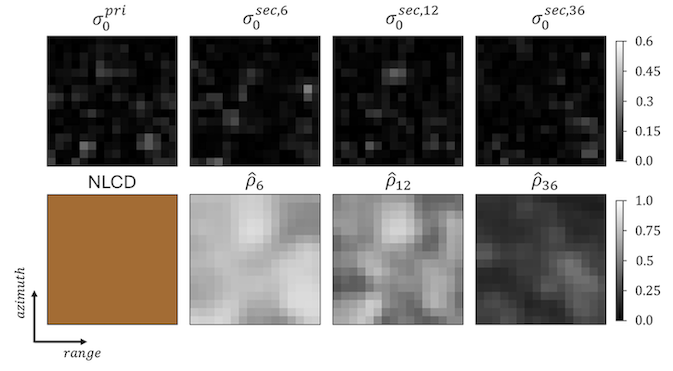}    
    \caption{Local-scale analysis over a cultivated region affected by temporal decorrelation (Crop II). The figure reports the temporal evolution of the input backscatter and the coherence predicted by the proposed framework across different temporal baselines.}
    \label{fig:crop16x16_cropland}
\end{figure}

\begin{table}[h]
\caption{Statistics of the backscatter coefficients. The table displays the mean value, standard deviation, Kolmogorov-Smirnov distance, and Kullback-Leibler divergence at 6-, 12-, and 36-day on crops I and II.} 
\label{tab:metrics_crops16x16} 
\centering
\renewcommand{\arraystretch}{1.5}
\begin{tabular}{c|c|cccc}
    \hline
    \multicolumn{6}{c}{\textbf{Statistics of the inputs}} \\
    \hline
    \textbf{Area} & \textbf{Metrics} & $\sigma_0^{pri}$ & $\sigma_0^{sec,6}$ & $\sigma_0^{sec,12}$ & $\sigma_0^{sec,36}$  \\
    \hline
    \multirow{4}{*}{\textbf{Crop I}} & Mean & $2.96$ & $2.78$ & $3.02$ & $2.77$ \\
    & Std. Dev. & $4.99$ & $4.84$ & $5.15$ & $5.49$ \\
    & K-S distance & - & $0.053$ & $0.053$ & $0.057$ \\ 
    & K-L divergence & - & $2.12$ & $1.49$ & $2.35$ \\
    \hline
    \multirow{4}{*}{\textbf{Crop II}} & Mean & $0.043$ & $0.036$ & $0.034$ & $0.030$ \\
    & Std. Dev. & $0.043$ & $0.042$ & $0.036$ & $0.032$ \\
    & K-S distance & - & $0.149$ & $0.194$ & $0.254$ \\ 
    & K-L divergence & - & $1.92$ & $1.66$ & $2.21$ \\
    \hline
\end{tabular}
\end{table}

The experimental results demonstrate that the proposed framework successfully learns statistical relationships associated with interferometric coherence levels at high spatial resolution using only detected SAR backscatter. When trained on 12-day Sentinel-1 data over Italy and evaluated on an independent 12-day pair over the San Francisco Bay area (Figure~\ref{fig:sanfrancisco_12days}), the model achieved an RMSE of 0.108, confirming its ability to generalize across geographically distinct regions and acquisition conditions.

An important question is whether the network learns meaningful relationships between paired SAR observations or simply exploits correlations between land cover and coherence. To investigate this aspect, an ablation study was conducted by retraining the network using only the primary acquisition as input (Figure~\ref{fig:sanfrancisco_12days}). In this configuration, the RMSE increased by 0.254 with respect to the dual-input model, demonstrating that the network effectively exploits the statistical relationship between the two acquisitions rather than relying solely on backscatter amplitude or scene semantics. This observation is particularly relevant because coherence is strongly influenced by temporal stability and scattering consistency, which cannot be fully inferred from a single image alone.

Further evidence of the model’s robustness is provided by the multitemporal analysis. Although the network was trained exclusively on 12-day interferometric pairs, it successfully generalized to 0-, 6-, and 36-day temporal baselines without additional retraining (Figures~\ref{fig:cropA_analysis}--\ref{fig:cropC_analysis}). The RMSE values remained relatively stable across all temporal configurations, indicating that the network captures statistical relationships associated with temporal decorrelation rather than memorizing a specific temporal baseline. This behavior suggests that the model learns generalized coherence-related representations linked to speckle evolution, texture consistency, and scattering stability across time.

Compared with the baseline coherence estimator proposed in \cite{MontiGuarnieri1997}, the proposed approach consistently achieved lower RMSE values across all evaluated areas (Table~\ref{tab:rmse_multitemp}). In addition to the quantitative improvements, the predicted coherence maps preserve finer spatial structures while exhibiting reduced noise and fewer multilooking artifacts. This is particularly visible in urban regions and heterogeneous scenes, where conventional intensity-based estimators suffer from spatial averaging effects.

As an additional consistency analysis, coherence was evaluated in the zero-baseline case by using identical acquisitions as input (Figure~\ref{fig:sanfrancisco_00days}). The predicted coherence values closely follow the expected SNR-limited coherence behavior, confirming that the model remains physically consistent even in the absence of temporal decorrelation. This experiment further supports that the network learns coherence-related statistical properties rather than merely reproducing empirical land-cover priors.

The experiments conducted on geographically distinct regions further support the robustness of the learned representation. Despite being trained exclusively on Sentinel-1 data over Italy, the framework successfully generalized to substantially different biome and scattering conditions, including the Atacama Desert, tropical rainforest regions in Brazil, and mountainous alpine environments in Switzerland (Figure~\ref{fig:vv_others_analysis}). In particular, the model accurately distinguished stable and decorrelating regions across very different environmental conditions, suggesting that the learned mapping is not limited to scene-specific land-cover priors. Minor degradation effects were mainly observed in regions characterized by strong geometric distortions and rapidly varying topographic conditions.

Additional experiments demonstrated that the proposed framework also generalizes across different acquisition configurations. The evaluation on VH-polarized Sentinel-1 data (Figure~\ref{fig:vh_analysis}) showed that the predicted coherence maps remain spatially consistent despite the absence of cross-polarized data during training. Although moderate performance degradation is observed with respect to the VV experiments, the obtained results indicate that the learned representation is not strictly tied to a specific polarization channel or radiometric distribution, but rather captures more general statistical relationships associated with coherence behavior.

Similarly, the large-baseline experiment over the Oberpfaffenhofen region demonstrated that the framework remains robust even under significantly different interferometric geometries (Figure~\ref{fig:op_vv_crop}). Despite the stronger geometric sensitivity and increased temporal decorrelation associated with the 453~m perpendicular baseline and 84-day temporal separation, the predicted coherence maps successfully reproduced the main spatial coherence structures across the scene. Nevertheless, the observed performance degradation suggests that geometric decorrelation effects cannot always be fully inferred from detected backscatter alone, particularly under highly non-nominal acquisition conditions.

The cross-frequency experiment conducted on ALOS L-band HH-polarized data further highlights the flexibility of the proposed methodology (Figure~\ref{fig:alos_Lband_hh}). Despite the substantial domain shift with respect to the Sentinel-1 C-band VV training data, the framework retained meaningful coherence prediction capability and reproduced the main decorrelation patterns across the scene. These results suggest that part of the learned representation may capture statistical relationships that remain informative across different frequency regimes. At the same time, the moderate increase in RMSE indicates that sensor- and frequency-specific fine-tuning would likely improve performance in operational cross-sensor scenarios.

To further investigate the learned representations at local scale, two homogeneous $16 \times 16$ pixel patches were analyzed: \textit{Crop~I}, corresponding to a stable urban area (Figure~\ref{fig:crop16x16_impervious}), and \textit{Crop~II}, corresponding to a cultivated region affected by temporal decorrelation (Figure~\ref{fig:crop16x16_cropland}). The statistical analysis reported in Table~\ref{tab:metrics_crops16x16} shows that decorrelating areas exhibit stronger divergence in speckle statistics over time. These observations indicate that local radiometric and textural variations provide meaningful cues for coherence regression and suggest that the network implicitly exploits both scene structure and temporal statistical consistency.
More generally, these observations suggest that the network exploits multiple complementary cues related to coherence behavior, including temporal consistency of speckle statistics, local structural similarity between acquisitions, texture stability, and radiometric divergence patterns associated with decorrelation processes. The combination of these factors likely enables the network to infer coherence-related representations beyond simple land-cover priors or intensity correlations alone.

Finally, the fine-tuning experiments demonstrate that adapting the statistical properties of SLC-derived backscatter enables successful transfer to geocoded GRD products. The proposed approach achieved coherent and spatially consistent predictions on the \textit{GEE-SAR Dataset} (Figures~\ref{fig:sanfrancisco_GEE_cropA}--\ref{fig:sanfrancisco_GEE_cropC}), despite the different radiometric and geometric characteristics of the data. These results confirm the feasibility of estimating coherence directly from globally available analysis-ready SAR archives, opening new perspectives for scalable interferometric applications on cloud computing platforms such as Google Earth Engine.

Although the obtained results demonstrate strong robustness across a broad range of acquisition conditions, some limitations should be acknowledged. The proposed framework does not explicitly reconstruct interferometric information and cannot fully model decorrelation mechanisms that are weakly observable from detected backscatter alone, particularly under extreme geometric acquisition conditions. In addition, the framework currently formulates coherence estimation as a deterministic regression problem and does not explicitly account for estimator uncertainty or confidence intervals. Finally, since the supervision is derived from coherence products estimated from coregistered SLC data, the proposed methodology may inherit part of the biases and artifacts associated with the reference coherence estimator.


\section{Conclusion}
\label{sec:conclusion}

This paper introduced a deep learning framework for interferometric coherence regression from detected SAR backscatter, without requiring full SLC interferometric processing and subpixel coregistration. The proposed methodology learns statistical and spatial relationships between paired backscatter observations, enabling accurate and high-resolution coherence regression across a broad range of acquisition conditions.

The experimental analysis demonstrated strong generalization capability across geographically distinct regions, multiple temporal baselines, polarization modes, interferometric baselines, and even different operative frequencies. Despite being trained exclusively on Sentinel-1 VV-polarized 12-day interferometric pairs, the framework successfully generalized to VH-polarized data, large-baseline interferometric configurations, and ALOS L-band acquisitions, while consistently outperforming conventional intensity-based coherence estimators.

Ablation experiments confirmed that the model effectively exploits the joint statistical information contained in both SAR acquisitions rather than relying solely on backscatter amplitude or scene semantics. In addition, the ability to reproduce the expected SNR-limited coherence behavior in zero-baseline configurations further supports the physical consistency of the learned representation.

The proposed framework was also successfully adapted to geocoded Sentinel-1 GRD products from Google Earth Engine, demonstrating the feasibility of generating meaningful coherence predictions directly from globally available analysis-ready SAR archives. This capability opens new perspectives for scalable interferometric analysis, rapid data screening, mission planning, and representation learning in SAR remote sensing applications.

Although the proposed methodology does not replace conventional interferometric processing and cannot fully model all decorrelation mechanisms from detected backscatter alone, the obtained results indicate that paired SAR intensity observations contain informative statistical cues related to coherence behavior. Future work will investigate uncertainty-aware formulations, cross-sensor adaptation strategies, and physics-informed learning approaches for improving robustness and interpretability across increasingly heterogeneous SAR acquisition scenarios.






\bibliographystyle{plainnat}
\bibliography{bibtex/bib/biblio.bib}
\vfill

\end{document}